\documentclass[a4paper,11pt]{article}
\pdfoutput=1
\usepackage{jheppub}
\usepackage{graphicx}
\usepackage{physics}
\usepackage{amsmath}
\usepackage{amsfonts}
\usepackage{textgreek}
\usepackage{amssymb}
\usepackage{slashed}
\usepackage{multirow}
\usepackage[colorlinks=true,linkcolor=blue]{hyperref}
\usepackage{bm}
\usepackage{empheq}
\usepackage[super]{nth}
\DeclareMathOperator\arctanh{arctanh}

\def\OO{\mathcal{O}}
\def\LL{\mathcal{L}}
\def\DD{\mathcal{D}}
\def\be{\begin{equation}}
\def\ee{\end{equation}}

\def\Eq#1{Eq.~\eqref{#1}}

\def\Tr{\,\mathrm{Tr}\,}
\def\SUW{SU_{\mathrm{w}}(2)}

\def\Gsphal{\Gamma_{\mathrm{sphal},s}}
\def\MS{\overline{\text{MS}}}
\def\alphas{\alpha_s}
\def\tE{t}
\def\QI{Q_{\mathrm{I}}}
\def\Qs{Q_{\mathrm{S}}}
\def\Qhalf{Q_{\mathrm{half}}}

\def\Ntau{N_\tE}
\def\Ns{N_s}
\def\betalatt{\beta_\text{latt}}
\def\tauf{{\tau_{\!_{\mathrm{F}}}}}
\def\tauft{{\tau_{\!_{\mathrm{F}3}}}}
\def\taufbar{{\overline{\tau}_{\!_{\mathrm{F}}}}}
\def\Tc{T_{\mathrm{c}}} 
\def\Tpc{T_{\mathrm{pc}}}
\def\tm{t_{\mathrm{m}}}
\def\ddd{\mathbf{d}}
\def\dd{\mathrm{d}}
\def\GammaE{\Gamma_{\mathrm{Eucl}}}
\def\ceven{c_{\mathrm{even}}}
\def\codd{c_{\mathrm{odd}}}

\title{The Sphaleron Rate from 4D Euclidean Lattices}
\author{Marc Barroso Mancha, Guy D.\ Moore}

\affiliation{Institut f\"ur Kernphysik, Technische Universit\"at Darmstadt\\
Schlossgartenstra{\ss}e 2, D-64289 Darmstadt, Germany}
\emailAdd{mbarroso@theorie.ikp.physik.tu-darmstadt.de,guy.moore@physik.tu-darmstadt.de}

\abstract{
We develop a new method to determine thermal activation rates, such as
for bubble nucleation, topology change, \textsl{etc.}, using
4-dimensional Euclidean methods.  This allows nonperturbative study
on the lattice.  We then investigate the strong sphaleron rate in
pure-glue QCD at temperatures between 1.3 $\Tc$ and 1000 $\Tc$, making
contact with previous results but extending them down close to the
critical temperature.  The extension to full QCD will be
straightforward.
Limitations of the proposal (the inability to compute a certain
dynamical prefactor, puzzling large-volume behavior, and the inability
to treat temperatures $T < 1.3 \Tc$) are also discussed.
}

\begin{document}
\maketitle
\section{Introduction}
\label{sec:intro}

Nonabelian gauge theories, including QCD and the $SU_{\mathrm{w}}(2)$
weak sector of the Standard Model, feature gauge groups which can have
topologically nontrivial configurations in 4 spacetime dimensions
\cite{Belavin:1975fg,tHooft:1976rip,tHooft:1976snw}.
Specifically, the spacetime integral of the contraction of the field
strength tensor%
\footnote{%
Our conventions are:  we use natural units, $\hbar = 1$ and $c=1$.
The Euclidean metric is positive and the
Minkowski metric is mostly-positive
$g_{\mu\nu} = \mathrm{Diag}[-1,+1,+1,+1]$;
the QCD/electroweak gauge fields are $G_\mu = G_\mu^a T^a$ and
$W_\mu = W_\mu^a T^a$ with $T^a$ the Hermitian
fundamental representation Lie algebra generators, normalized to
$\Tr T^a T^b = \frac{\delta_{ab}}{2}$;
the fields take their geometrical normalization such that the gauge
coupling does not appear in the covariant derivative
$D_\mu = \partial_\mu -i G_\mu$ but as a denominator in the gauge
action $\mathcal{L} = \ldots - \Tr G_{\mu\nu} G^{\mu\nu}/2g^2$.
}
$G_{\mu\nu}$ with the dual field strength tensor
$\tilde{G}^{\mu\nu} \equiv \epsilon^{\mu\nu\alpha\beta}
G_{\alpha\beta}/2$, satisfies:
\begin{equation}
  \label{NI}
  q(x) \equiv \frac{\Tr G_{\mu\nu} \tilde{G}^{\mu\nu}}
  {16\pi^2} \,, \qquad
  \QI = \int \dd^4 x \; q(x) \; \in \mathbb{Z} \,.
\end{equation}
That is, the integral of $q(x)$ the ``topological charge density''
always returns an integer, provided the integral is carried out
over a compact domain without boundaries or with boundaries where the
field strength goes to zero.  (For the case of the $\SUW$ sector,
replace $G_{\mu\nu} \to W_{\mu\nu}$.)

These topological structures are important because they lead to
anomalies in axial quark number currents:  according to the
Adler-Bell-Jackiw anomaly
\cite{Adler:1969gk,Bell:1969ts}, each fundamental-representation fermionic
species suffers a nonconservation of its axial charge in the amount
\be
\label{J5}
\partial_\mu J^\mu_5(x) = \ldots + 2 q(x) \,,
\qquad
\Delta Q_5 = 2 \QI \,.
\ee
Here $J^\mu_5$ is the N\"other current associated with axial quark
number and $Q_5$ is the associated axial number.
In the electroweak sector the effect is dramatic:  because only
left-handed species couple to $\SUW$, both baryon number and lepton
number are violated,
\cite{tHooft:1976snw}:
$\partial_\mu J^\mu_B = N_g \frac{\Tr W_{\mu\nu} \tilde{W}^{\mu\nu}}{16\pi^2}$
with $N_g=3$ the number of generations.
Topological transitions in real time can be stimulated if the fields
can go over the so-called sphaleron barrier
\cite{Klinkhamer:1984di}, and this process becomes efficient at high
temperatures \cite{Kuzmin:1985mm} and may explain the baryon number of
the Universe (see for instance \cite{Rubakov:1996vz,Cline:2006ts} and
references therein).

In Quantum Chromodynamics topological transitions in real time are also
relevant for interesting high-temperature dynamics.
They are responsible for the equilibration between right- and
left-handed quark number in electroweak baryogenesis scenarios
\cite{McLerran:1990de,Giudice:1993bb}.
In the context of relativistic heavy ion physics, noncentral
collisions contain intense magnetic fields, which can combine with
axial quark number imbalances to provide interesting new physics
signals -- the so-called chiral magnetic effect
\cite{Fukushima:2008xe}.
This proposal has inspired a great deal of work, including a deeper
appreciation of the hydrodynamics of systems with axial quark number
\cite{Son:2009tf,Stephanov:2012ki}.

Such chiral dynamics in QCD are very dependent on the decay rate of
chiral imbalances.
Close to equilibrium, a net axial charge abundance is described by a chemical
potential for axial quark number $\mu_A$; at leading order and in each
quark species, the axial number density $n_A$ is related to $\mu_A$
through \cite{Giudice:1993bb}
\begin{equation}
  \label{suscept}
  n_A = \mu_A T^2 \,.
\end{equation}
The coefficient is determined by the quark number susceptibility,
which is known \cite{Bazavov:2013uja} to be within $10\%$ of the
leading-order value already at $T = 2 \,\Tpc$ with $\Tpc$ the
pseudocritical temperature, $\Tpc \simeq 155$ MeV
\cite{Borsanyi:2013bia,HotQCD:2014kol}.
The axial chemical potential biases thermal transitions and thereby
acts to create a net $G_{\mu\nu} \tilde{G}^{\mu\nu}$ value which
drives down the axial number.  The Minkowski time $\tm$ derivative of
$n_A$ due to this thermal relaxation effect is:
\begin{equation}
  \label{dndt}
  \frac{\dd n_A}{\dd \tm} = - \frac{\sum_f \mu_{A,f}}{T} \Gsphal
  = -n_A \frac{n_f \Gsphal}{T^3} \,,
\end{equation}
where $n_f$ is the number of light fermionic species and $\Gsphal$,
the so-called \textsl{strong sphaleron rate},
describes the equilibrium diffusion rate of topology due to random,
real-time topological processes at temperature $T$:
\begin{equation}
  \label{Gsphaldef}
  \Gsphal \equiv \frac{1}{Vt_{\mathrm{max}}} \left\langle \left(
  \int_0^{t_{\mathrm{max}}} \dd \tm \int_V \dd^3 x \: q(x,\tm) \right)^2 \right\rangle
      = \int \dd^4 x \langle q(x) q(0) \rangle \,.
\end{equation}
That is, $\Gsphal$ is the mean-squared topology per unit Minkowski
4-volume.%
\footnote{Some older sources define the sphaleron rate such that
$dn_A/d\tm = - \sum_{f,s} \mu_{A,fs} \Gamma/T$ where the sum also runs over
spins $s$ -- that is, the sum is over all species which would be created
by a sphaleron.  This definition is half as large as the one we use,
that is, $\Gamma = \Gsphal/2$. \label{footfactor2}}
The definition is similar to the topological
susceptibility $\chi_{\mathrm{top}}$, except that
$\chi_{\mathrm{top}}$ refers to correlators in Euclidean spacetime,
while the time appearing in \Eq{Gsphaldef} is real Minkowski time.  As
we will discuss below, these two rates have different physical
interpretations and relevance and at high temperatures they strongly
differ from each other.

In order to understand the dynamics of chiral imbalances, it is
imperative to have at least a decent determination of the
sphaleron rate $\Gsphal$.  If it is large, chiral imbalances relax
quickly and are unlikely to lead to interesting dynamics.  On the
other hand, if it is small, it may be a good approximation to treat
axial quark number as approximately conserved, at least for light
species.

Unfortunately, the state of the art in determining $\Gsphal$ is not
very advanced.  The equivalent $\SUW$ rate is relatively well
understood.  The relevant dynamics involves soft (wave number small
compared to $T$) gauge fields which should behave like classical
fields \cite{Grigoriev:1988bd}, though the relevant dynamics are
subtle \cite{Bodeker:1995pp,Arnold:1996dy,Bodeker:1999gx}.
Building on early work by Ambj{\o}rn et al
\cite{Ambjorn:1990pu},
a fairly complete picture for the sphaleron rate in the $\SUW$ sector
emerged by the end of the 1990s
\cite{Ambjorn:1997jz,Moore:1997sn,Arnold:1997yb,Bodeker:1998hm,
  Moore:1998zk,Bodeker:1999gx,Arnold:1999uy}.
This success is based on the fact that the $\SUW$ sector of the
Standard Model has a rather small gauge coupling,
$g^2 N/4\pi \simeq 1/15$, which ensures that the relevant dynamics is
quite infrared and the classical-field approximation is well under
control.  In comparison, even at electroweak temperatures
$T \sim 1000 \, \Tpc$, the strong coupling is of order
$g^2 N/4\pi \sim 0.3$, and the classical-field approximation
is not very reliable.
There are only two attempts to determine the strong sphaleron rate
nonperturbatively using lattice techniques
\cite{Moore:1997im,Moore:2010jd}, and the latter concludes that the
results are no better than factor-of-2 estimates even
at $T = 100\,\mathrm{GeV} \sim 700 \,\Tpc$.  The rate below this
temperature is not even estimated.

We need a new approach to compute the strong sphaleron rate -- one
which does not rely on an approximation of QCD dynamics in terms of
3+1 dimensional classical fields.  The approach must be
nonperturbative, because the temperatures achieved in the early
stages of heavy ion collisions are at most a few times higher than
$\Tpc$, where the theory is fully nonperturbative.
Therefore the approach must involve our only rigorous tool for
understanding QCD in this regime, which is 4-dimensional Euclidean
lattice gauge theory.

One approach proposed recently
\cite{Altenkort:2020axj,Altenkort:2021jbk} is to measure $q(x)\,q(0)$
correlators as a function of Euclidean time and to attempt an analytic
continuation to obtain the real-time spectral function, whose
low-frequency limit determines $\Gsphal$.  This is an interesting
approach, but it is very challenging, particularly because of our
ignorance of the expected low-frequency behavior of the spectral
function and the challenge in reconstructing nontrivial low-frequency
structures in such analytical continuations.

We will provide a completely new approach for estimating $\Gsphal$
from Euclidean lattice path-integral data.
Our approach is based on the picture of thermal activation developed
by Langer \cite{Langer:1969bc} and further developed by Affleck
\cite{Affleck:1980ac} and Linde \cite{Linde:1980tt,Linde:1981zj}.
The next section (Sec.~\ref{sec:activation}) will review this
picture.  Thermal activation is controlled by Euclidean configurations
which are balanced at a saddlepoint between two (topological) minima,
with a rate determined by the prevalence of saddlepoint configurations
times the real-time rate that the fields move through such saddles.
In Subsection \ref{sec:fluct} we show that this rate is related to the
size of fluctuations in the Euclidean direction, such that different
Euclidean-time slices lie on different sides of the saddle.
This is something which can be measured purely within Euclidean
simulations.  We formulate how to do this specifically for the QCD
sphaleron rate (that is, $SU(3)$ gauge theory with or without
fermions) in Section \ref{sec:QCD}, and show how to really carry out
such calculations in Section \ref{sec:strong-sph-rate}.

With the method in hand, we make the usual checks for lattice spacing,
lattice volume, and gradient-flow depth dependence, within pure-glue
QCD ($SU(3)$ Yang-Mills theory) in Section \ref{sec:tests}.  Having
passed all tests, we explore what range of temperatures we can
investigate in Section \ref{sec:instanton}, finding that the method
breaks down below about $T = 1.3 \, \Tc$.  Our final results for
temperatures above this threshold are presented in Section
\ref{sec:results}, followed by a discussion and outlook.

\section{Thermal activation:  a Review}
\label{sec:activation}


The theory of thermal activation was pioneered by Langer
\cite{Langer:1969bc} and put into field-theoretical language by
Affleck \cite{Affleck:1980ac} and by Linde \cite{Linde:1981zj}.
The canonical example is the problem of bubble nucleation in a system
with a stable and a metastable vacuum.  At low or zero temperatures,
this is controlled by a 4-dimensional bubble configuration, as argued
by Coleman and Callan
\cite{Coleman:1977py,Callan:1977pt}.
At higher temperatures (roughly, when $1/T$ is of order or smaller than
the diameter of the would-be 4D bubble), the relevant solution is
instead a bubble which is constant in the (periodic,
$\beta=1/T$-extent, Euclidean) time direction and varies as a function
of spatial radius.  The field theoretical justification is worked out
by Affleck \cite{Affleck:1980ac} and by Linde
\cite{Linde:1980tt,Linde:1981zj}.

However, for our purposes this example is confusing and involves
extraneous details.  Instead, we will outline the key ideas, and
motivate the approach we will take, with a much simpler toy example,
proposed by Arnold and McLerran.

\subsection{Simple example:  pendulum under gravity}
\label{secpendulum}

Consider a rigid pendulum in a gravitational potential.  There is one
degree of freedom $\varphi$, and the Lagrangian is
\begin{equation}
  \label{Lpendulum}
  \LL(\varphi,\dot\varphi) = \frac{m}{2} \dot\varphi^2
  -V_0 \left(1 - \cos\varphi \right) \,.
\end{equation}
Arnold and McLerran used this problem to illustrate the difference
between instantons and sphalerons and the physical role each plays
\cite{Arnold:1987zg}, and we are guided by their considerations.

First, note that the energy spacing between the low-lying quantum
excitations for this system is $\omega_0 = \sqrt{V_0/m}$, whereas the
energy required to spin the pendulum over the top of its potential is
$2V_0$.  We will consider the case
$2V_0 \gg \omega_0$ and will consider temperatures
$T < \omega_0$, where the vacuum dominates, and
$2V_0 \gg T \gg \omega_0$, where a range of thermal states participate
but states with $E>2V_0$ are still rare.
To make the thermal ensemble make sense and the real-time behavior be
more similar to a hot, chaotic system, we will couple $\varphi$ to a
thermal bath of oscillators.  We assume that this bath causes a negligible
modification of the above Lagrangian, but that it leads to noisy,
dissipative real-time dynamics.  We will not specify these couplings
in detail, since this is only an illustrative example for us and not
our main item of study.

Let us understand the distinction between instantons and
sphalerons, and the role of each.  First the instanton.  The classical
Lagrangian does not actually specify our problem completely, because
the values $\varphi = 0$ and $\varphi = 2\pi$ (or any pair of values
separated by $2\pi$) are physically equivalent, and the wave function
must be periodic up to a phase,
$\Psi(\varphi+2\pi) = \Psi(\varphi) e^{i\theta}$.
Here $\theta$ is a parameter, analogous to the $\Theta$-parameter of
QCD, which only plays a role in the quantum theory.
Instantons are configurations which answer the question:

\begin{verse}
  \textsl{Instantons:} how strongly does the free energy depend on
  $\theta$?
\end{verse}

We can address the $\theta$-dependence of the free energy by
considering its calculation with the Euclidean path integral.
Naively we expect in the Euclidean path integral that $\varphi$ is
periodic, $\varphi(t=0) = \varphi(t=\beta)$, but in fact we only
require periodicity modulo $2\pi$.  However, configurations with
$\varphi(t=\beta) = \varphi(t=0) + 2\pi n$ should be weighted
with a factor $e^{in\theta}$, such that the thermal partition function
is:
\begin{equation}
  \label{ZT}
  Z(T) = \sum_{n \in \mathcal{Z}} e^{in\theta} \int_{-\pi}^\pi \dd\varphi_0
  \langle \varphi_0 | e^{-\beta H} |\varphi_0+2\pi n \rangle
  = \sum_n e^{in\theta} Z_n(T) \,.
\end{equation}
The $\varphi_0$ integral is a trace over all states, with the $n$-sum
reflecting the fact that these states are $2\pi$-periodic but allowing
us to write a path integral in terms of states which do not have this
periodicity.
We name the configurations contributing to $Z_n$ ($n\neq 0$)
$n$-instantons.

\begin{figure}[htb]
      \centering
    \includegraphics[trim={0 0cm 0 0cm},clip,width=1.0\textwidth]{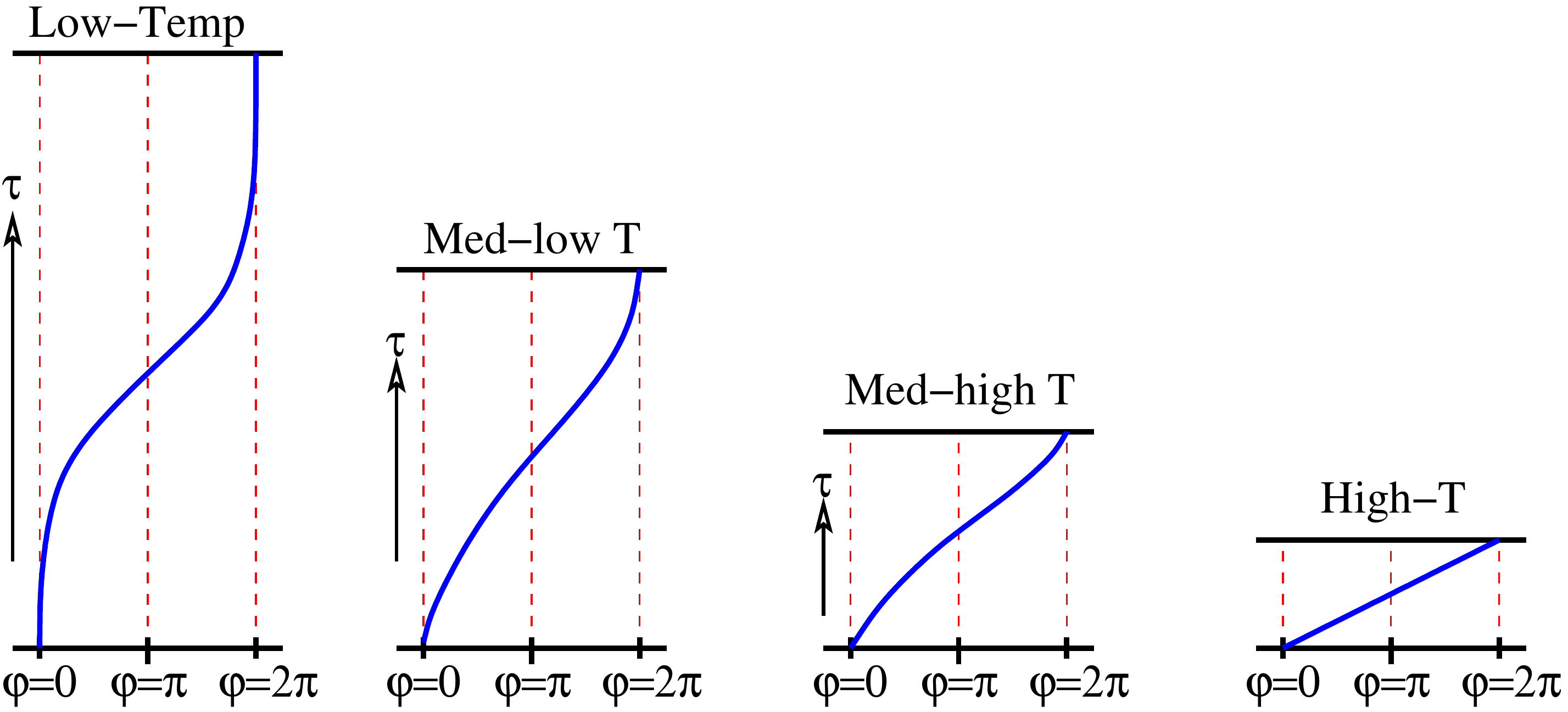}
  \caption{\label{fig:hiTinstanton}
    Instantons as a function of temperature.
    At low temperatures, $\varphi$ transitions from $0$ to $2\pi$
    along a smooth characteristic curve.  With increasing $T$, as the
    range of $t$-periodicity is narrowed, the instanton must
    ``hurry'' between $\varphi=0$ and $\varphi=2\pi$, increasing the
    associated action.}
\end{figure}

At low temperature $Z(T) = e^{-\beta E_0}$ is dominated by the vacuum
energy and the instantons determine how much the vacuum-energy depends
on the boundary condition.  For $\theta=0$ all $Z_n$ add and $E_0$ is
minimized, as we expect for a periodic boundary, whereas for
$\theta=\pi$ $E_0$ takes its maximal value because the vacuum wave
function is forced to have a node at $\varphi=\pi$.
At higher temperature the behavior is more complex.
Even-parity (at $\varphi=0$) states have the lowest energy for
$\theta=0$, but odd-parity states have the highest energy for
$\theta=0$ and the lowest energy for $\theta=\pi$.  At temperatures
well above $T=\omega_0$, the free energy is determined by a mixture of these
states and the effects of $\theta$ tend to cancel out.  In the
Euclidean path integral, this happens because the characteristic
$t$-width (Euclidean time width) over which $\varphi$ varies in an
instanton is $t \sim 1/\omega_0$.  When $\beta \omega < 1$, the
$\varphi$ value must change very rapidly in order to complete an
instanton, such that
$S_{\mathrm{inst}} \geq \int_0^\beta \dd t \: m(\partial_t \varphi)^2/2 \geq 2\pi^2 mT$.
The instanton action rises linearly at large $T$,
rendering $\theta$ irrelevant.  This point is illustrated in Figure
\ref{fig:hiTinstanton}.

Now consider the same system, but distinguishing between $\varphi=0$
and $\varphi=2\pi$ as physically distinct configurations.  In this
case, we can consider a setup where the initial conditions are close
to thermal, but with $\varphi$ always close to the minimum at 0, and
ask the completely different question:

\begin{verse}
  \textsl{Sphalerons:}  How fast does
  $\varphi$ diffuse over the range of available $\varphi=2\pi n$
  minima over very long periods of Minkowski time?
\end{verse}

\noindent
We can define this rigorously in terms of the long-time diffusion
of $\varphi$'s value, distinguishing between $2\pi$-copies, as
$\Gamma = \lim_{\tm \to \infty} \frac{1}{(2\pi)^2 \tm} \langle
(\varphi(\tm)-\varphi(0))^2\rangle$.
This is controlled by transitions over $\varphi=\pi$ and its periodic
copies because the
potential has a maximum there and this represents the principal
barrier to $\varphi$ evolution.%
\footnote{The definition of $\Gamma$ described in footnote \ref{footfactor2} is
the rate of \textsl{forward} transitions, $\varphi < \pi$
transitioning to $\varphi > \pi$.  The analogous rate is the relevant
one for the bubble nucleation problem and is therefore the one
considered by Affleck \cite{Affleck:1980ac}.  Again, the rate we
consider is for transitions in both directions and is precisely twice
as large.  This factor emerges when computing
$\langle |\dd \varphi / \dd \tm |\rangle$, which is the mean rate of
crossing in \textsl{either} direction.  To compute the mean rate of
forward crossings only, one should average this mean field velocity
only over the forward-moving cases.}

\begin{figure}[tbh]
  \includegraphics[trim={0 0cm 0
      0cm},clip,width=0.45\textwidth]{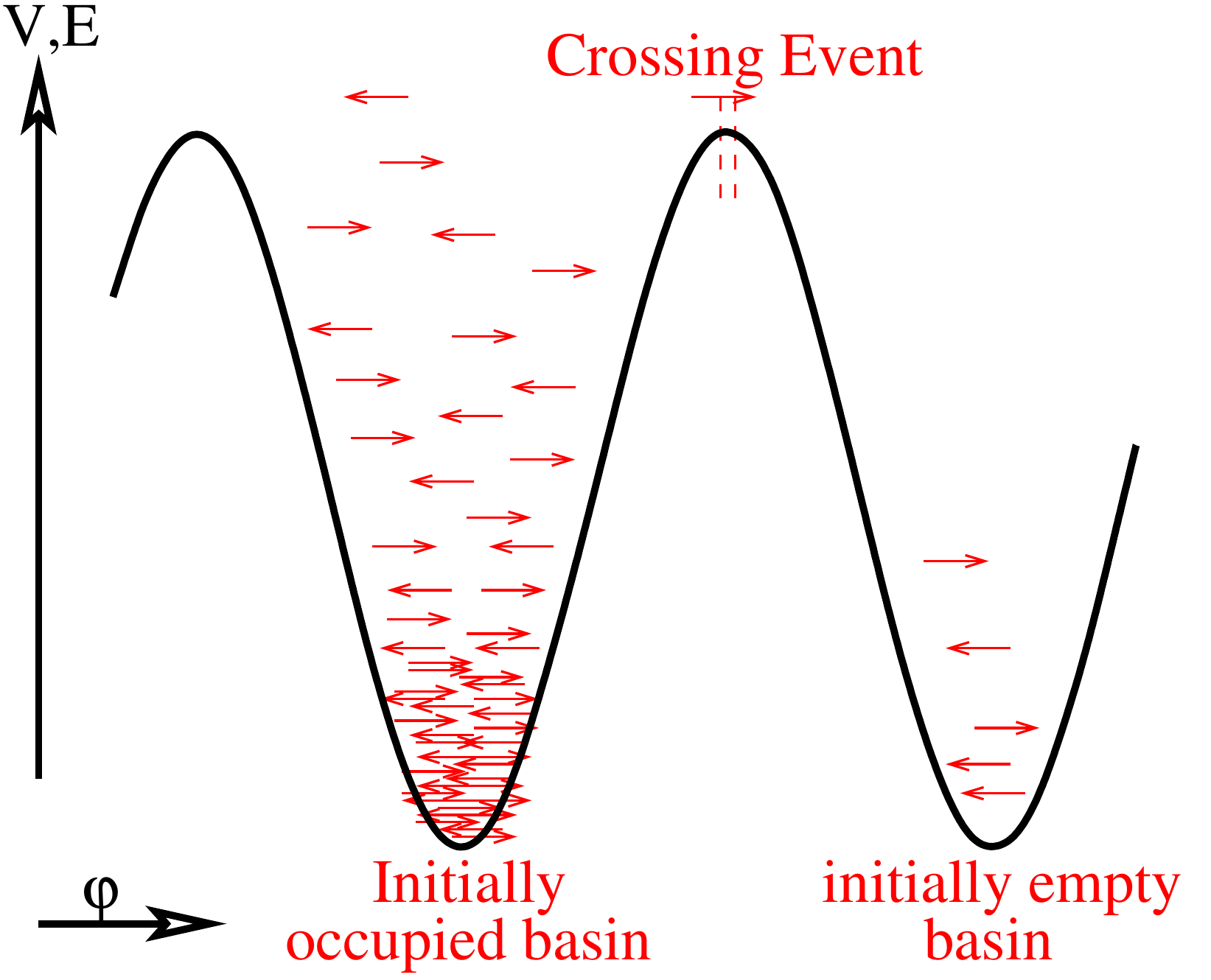} \hfill
  \includegraphics[trim={0 0cm 0
      0cm},clip,width=0.45\textwidth]{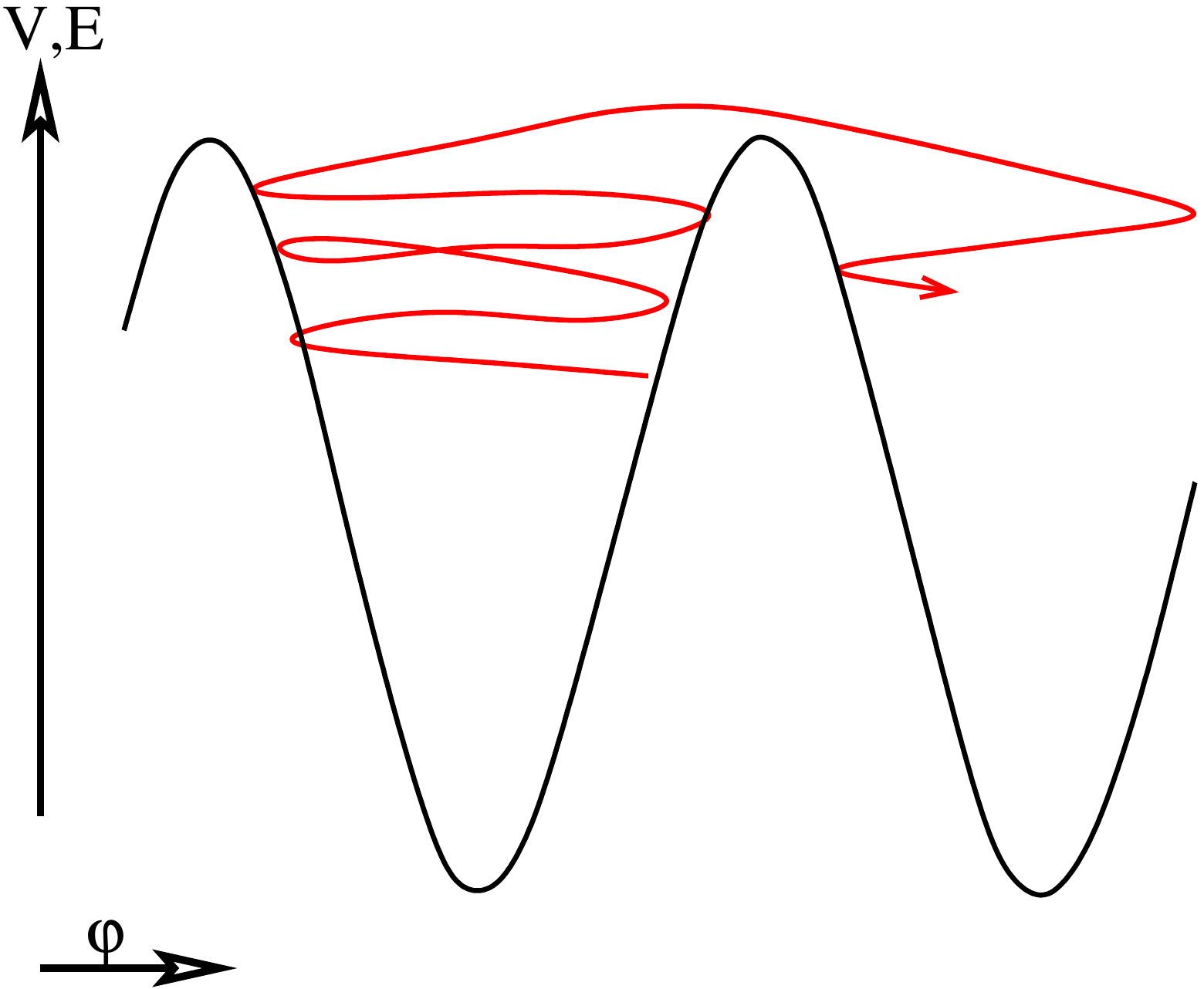}
  \caption{\label{fig:varphicartoon}
    Cartoon of thermal activation in the pendulum toy model.
    Left: when one basin starts filled and one starts empty,
    the rate to move across is the flux across
    the top, which is the probability to be near the top times the
    rate of crossing.
    Right:  for typical damping, an evolution bounces in one basin,
    gains enough energy to cross, and then bounces in the next basin.
  }
\end{figure}

Intuitively, the sphaleron rate is the product of three factors (see figure \ref{fig:varphicartoon}):
\begin{enumerate}
\item
  \label{dpdphi}
  How likely is $\varphi$ to take a value near the top of the barrier?
  That is, what is $\dd P(\varphi)/\dd\varphi$ at $\varphi=\pi$,
  where $\dd P(\varphi)/\dd\varphi$ is the probability density to find
  $\varphi$ in some narrow range about a specific value.
  In the toy model we expect $\dd P/\dd\varphi \sim \exp(-2V_0/T)$.
\item
  \label{dphidt}
  How fast does $\varphi$ change values when it is at the top of the
  barrier?  For our problem this is
  $\langle | \dd\varphi/\dd\tm | \rangle$.
  Note that a rescaling of the variable $\varphi$ changes both
  \ref{dpdphi} and \ref{dphidt}, but the change cancels such that the
  product of the two terms is rescaling invariant.
  In the toy model we expect $m\dot\varphi^2 \sim T$, so
  $\langle |\dd\varphi/\dd\tm| \rangle \sim \sqrt{T/m}$.
\item
  \label{dynprefact}
  How correlated are successive crossings of $\varphi=\pi$?
  This is a pure number, usually close to 1, which we will call
  $\ddd$.
\end{enumerate}
The total transition rate is then
\begin{equation}
  \label{totalrate}
  \Gamma = \left. \frac{\dd P(\varphi)}{\dd\varphi}\right|_{\varphi=\pi}
  \times \left\langle \left| \frac{\dd\varphi}{\dd\tm} \right| \right\rangle
  \times \ddd \,.
\end{equation}

\begin{figure}[bth]
  \includegraphics[width=0.4\textwidth]{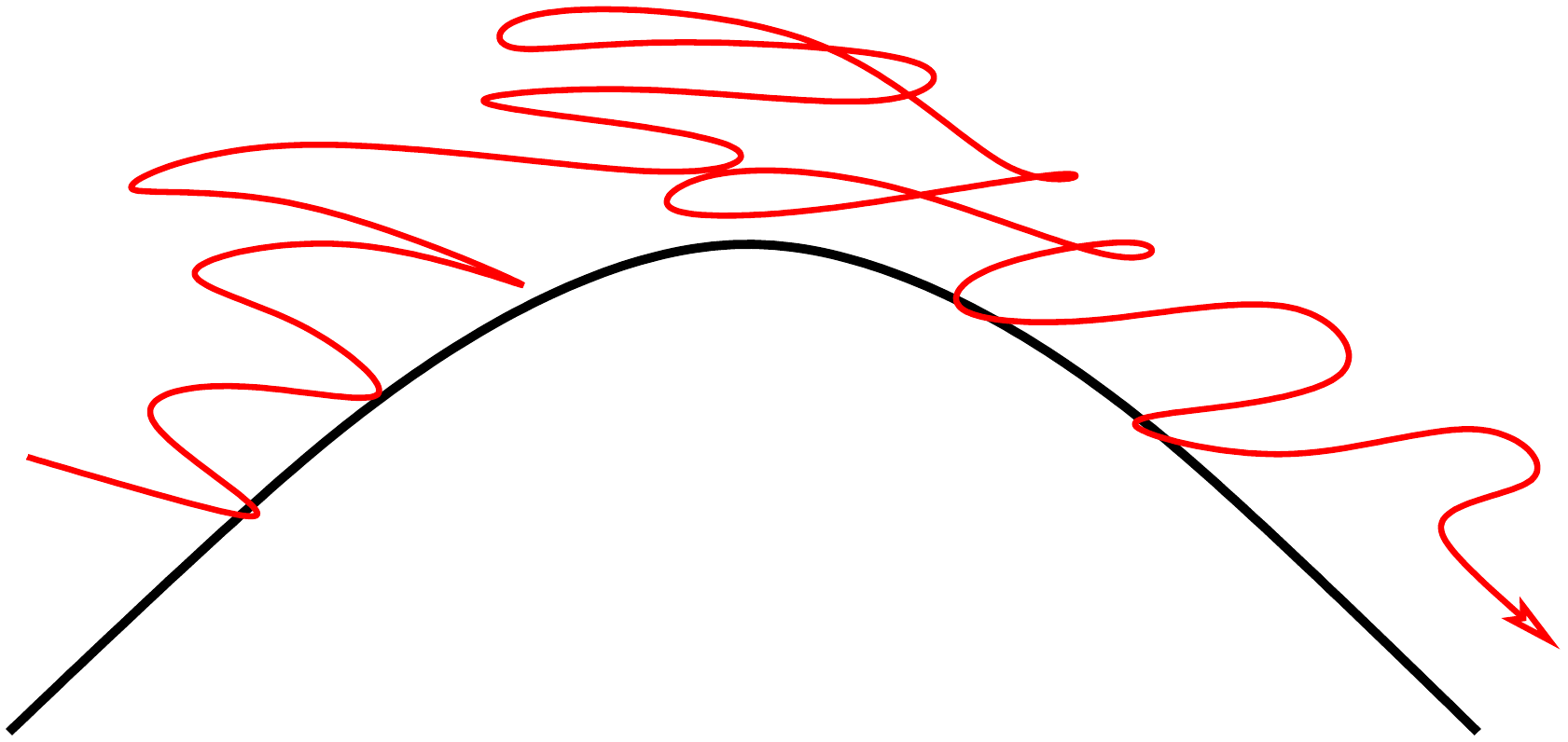} \hfill
  \includegraphics[width=0.5\textwidth]{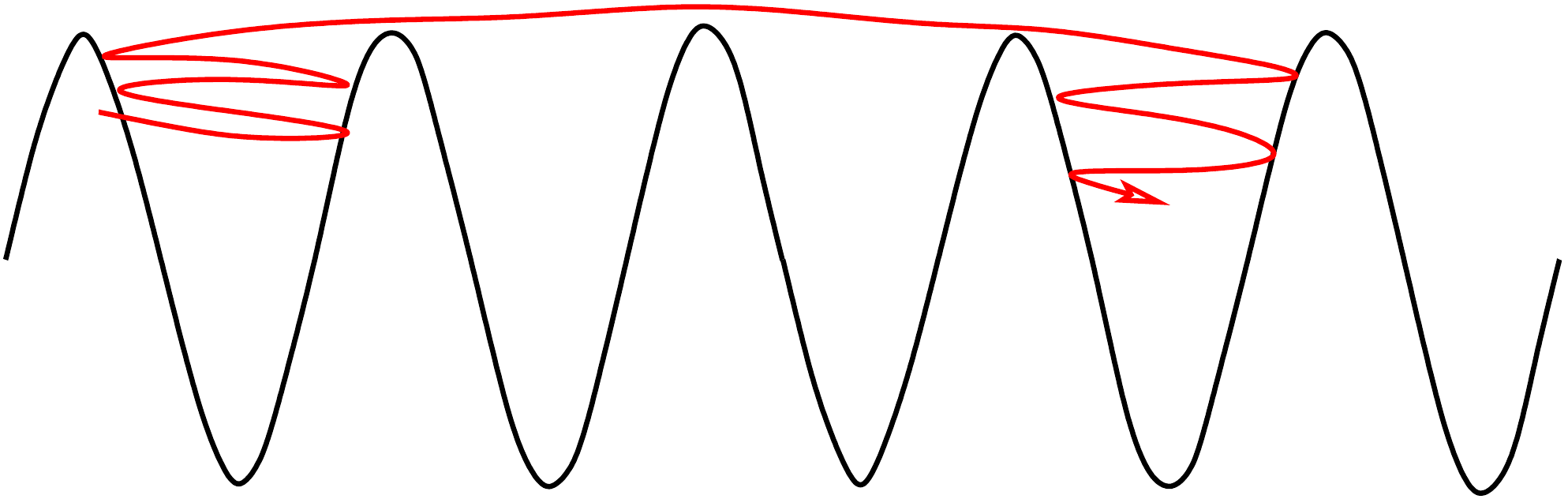}
  \caption{\label{fig:prefact}
    Cartoon illustrating the dynamical prefactor $\ddd$.
    For extremely strong damping (left), the trajectory can cross the barrier
    several times in succession, leading to only one (or zero) true
    crossings.  For underdamping (right), it can cross several barriers in
    succession.}
\end{figure}

The first two points are intuitively clear, but are explained in
proper detail in the seminal paper of Affleck.
The last point appears to have been first noticed by Arnold and
McLerran \cite{Arnold:1987mh}, and is illustrated in Figure
\ref{fig:prefact}.  By very weak damping, the system can cross several
barriers in succession.  Crossings are positively correlated and the
number of individual crossings underestimates the \textsl{square} of
the change in $\varphi/2\pi$.  In fact, in the absence of damping as
considered by Arnold and McLerran \cite{Arnold:1987zg}, $\ddd$
is infinite and the mean-squared $\varphi$ value rises quadratically,
rather than linearly, with time.  But for very strong overdamping,
several crossings may occur in a single ``macroscopic'' crossing event
-- as appears to occur for the electroweak sphaleron rate
\cite{Arnold:1996dy,Bodeker:1999gx}.  In this case, the true rate is
smaller than the count of individual crossings.
In a broad intermediate range of damping strengths, the $\dot\varphi$
direction does not change as it crosses the top of the barrier (say,
the region where $V_{\mathrm{max}}-V < T$), but there is enough damping
by the time it reaches the \textsl{next} barrier that it cannot cross.
In this case we expect $\ddd \simeq 1$.

We should emphasize that the technique we develop here \textsl{cannot}
determine $\ddd$.  Instead, we will concentrate on
determining the product of the first two factors, and will assume that
$\ddd$ is close to 1.  This is a weakness of our
approach, and it is not clear to us how it can be overcome.

In a more general context, the exact location of and distance from the
barrier may be harder to quantify.  Therefore, a successful thermal
activation calculation typically requires a few things:
\begin{enumerate}
\item
  Identify the ``separatrix'' which divides configurations into those
  closer to one minimum and those closer to another.
\item
  Identify some measure of distance $\varphi$ from the separatrix.
\item
  Use a Euclidean path integral to find the probability to be within
  some narrow tolerance $|\varphi| < \Delta$ of the separatrix.
\item
  Determine the mean velocity of the $\varphi$ variable when it is at
  or very near the separatrix.
\end{enumerate}
We will show how to combine the last 3 steps into a single step, which
does not require determining the value of $\varphi$.
  
\subsection{Crossing speed and Euclidean fluctuations}
\label{sec:fluct}

How do we estimate $\langle |\dd\varphi/\dd t|\rangle$ (or its equivalent
in more complex problems) using only Euclidean data?
The answer lies in the size of the fluctuations in $\varphi(t)$ as a
function of the Euclidean time $t$.
The amplitude of these fluctuations is proportional to the rate at
which $\varphi$ changes in real time.
After all, the real-time evolution can be determined by analytical
continuation from the Euclidean-time dependence, as calculated in
detail by Arnold and McLerran \cite{Arnold:1987zg}.
Consider again our toy example, with Lagrange function given in
\Eq{Lpendulum}.  The Hamiltonian is
\begin{equation}
  \label{Hpendulum}
  H = \frac{m}{2} \dot\varphi^2 + V_0 ( 1 - \cos(\varphi) )
\end{equation}
and classically, the thermal distribution of $\dot\varphi$ values is
given by equipartition,
\begin{align}
  \nonumber
  P(\dot\varphi) & \propto \exp( -m \dot \varphi^2 / 2T )
  \\
  \label{equipartition}
  \left\langle \left| \frac{\dd \varphi}{\dd \tm}
  \right| \right\rangle & =
  \frac{\int_0^\infty \dd\dot\varphi \; \dot\varphi
    e^{-m\dot\varphi^2/2T}}{\int_0^\infty \dd\dot\varphi \;
    e^{-m\dot\varphi^2/2T}} = 
  \\ \nonumber
  & = \sqrt{\frac{2T}{\pi m}} \,.
\end{align}

For comparison, in Euclidean space, the dominant $t$-dependence in
$\varphi(t)$ arises from fluctuations with Matsubara frequency
$\omega=2\pi T$:
\begin{equation}
  \label{Matsubara2}
  \varphi(t) \simeq \varphi_0 + c \, \cos(2\pi tT)
  + s \, \sin(2\pi tT)
\end{equation}
with $\varphi_0$ the average of $\varphi(t)$ over $t$ values and
$c,s$ coefficients indicating two independent $t$-dependent
fluctuations with this frequency.  The action associated with these
fluctuations is approximately
\begin{equation}
\label{Smatsubara}
  S(c,s) = \int_0^\beta \!\!\! \dd t \; \frac{m}{2} (\partial_t \varphi)^2
  = \int_0^\beta \!\! \dd t\; 2\pi^2 mT^2
  \left( c^2 \cos^2(2\pi tT) + s^2 \sin^2(2\pi tT) \right)
  =\pi^2 mT (c^2 + s^2) \,.
\end{equation}
We can characterize $\varphi$'s $t$ dependence by looking at the RMS
difference between $\varphi(t=\beta/2)$ and $\varphi(t=0)$,
which equals $2c$:
\begin{equation}
  \label{MeansqMatsubara}
  \langle (\varphi(\beta/2)-\varphi(0))^2 \rangle = 
  \langle (2c)^2 \rangle =
  \frac{\int_{-\infty}^\infty \dd c (2c)^2 e^{-\pi^2 mT c^2}}
       {\int_{-\infty}^\infty \dd c e^{-\pi^2 mT c^2}}
       = \frac{2}{\pi^2 mT} \; \Rightarrow \;
       \sqrt{\langle (2c)^2 \rangle} = \sqrt{\frac{2}{\pi^2 mT}}
\end{equation}
which is smaller than
$\langle |\dd \varphi/ \dd \tm | \rangle$
by a factor of $1/T\sqrt{\pi}$.  So
the side-to-side fluctuations in $\varphi$ are smaller than
$|\dd\varphi/\dd t|$ by a factor of $1/T$, which one could guess on
dimensional grounds, and an order-1 numerical factor.

Is this relation special?  We claim it is quite general -- the same
Hamiltonian controls real-time evolution and Euclidean-time
fluctuations, since the Hamiltonian $H$ equals the Euclidean-time
Lagrangian $\LL_{\mathrm{E}}$.

To determine the numerical factor precisely, we have to decide
precisely what question we want to ask about the Euclidean $\varphi$
fluctuations.  The point $\varphi=\pi$ separates values closer to the
minimum at 0 and the values closer to the minimum at $2\pi$, so we
follow standard usage and call it a separatrix.
In the following, it will be useful to ask how often
$\varphi(t=0)$ and $\varphi(t=\beta/2)$ lie on opposite sides of this
separatrix.  In particular, we expect on the above reasoning that
the first two terms in \Eq{totalrate} can be recast as:
\begin{equation}
  \label{totalrate2}
  \left.  \frac{\dd P(\varphi)}{\dd \varphi} \right|_{\varphi=\pi}
  \times \left\langle \left| \frac{\dd \varphi}{\dd \tm}
  \right| \right\rangle
  = \frac{\mbox{order-1 coeff}}{T} \times
  2\, \mathrm{Prob}\Big(\varphi(t=0)<\pi<\varphi(t=\beta/2)\Big) \,.
\end{equation}
Here $2\,\mathrm{Prob}(...)$ is the chance that $\varphi$ is on opposite
sides of $\pi$ at the Euclidean times $t=0$ and $t=\beta/2$; the
factor of $2$ accounts for $\varphi(t=0)>\pi>\varphi(t=\beta/2)$.

To determine the order-1 coefficient, we must repeat the quick
calculation of $(2c)_{\mathrm{RMS}}$ above with more care.
We do so in Appendix \ref{sec:thm-euclidean}, accounting for all Matsubara
modes, finding that the real-time rate is $2T$ times the probability
for $\varphi(t=0)$ and $\varphi(t=\beta/2)$ to be on opposite sides of
the separatrix.%
\footnote{Here is a quick way to see why.  Compare \Eq{equipartition}
to \Eq{MeansqMatsubara}.  The latter was computed with only the lowest
Matsubara frequency, but all odd frequencies contribute, multiplying
it by a factor of $\sqrt{\sum_{n=1,3,\ldots} n^{-2} = \pi^2/8}$.
Also, \Eq{MeansqMatsubara} is based on an RMS average, but we should
compute the mean absolute value as used in \Eq{equipartition}, which
is smaller by $\sqrt{2/\pi}$ for Gaussian distributions.  The two
effects correct \Eq{MeansqMatsubara} to
$\sqrt{2/(\pi^2 mT) \times \pi^2/8 \times 2/\pi} = (1/2) \sqrt{2/\pi mT}$
which must be multiplied by $2T$ to get \Eq{equipartition}.}
The appendix also shows how to include the effects of
gradient flow, which we will need later.

The advantage of this formulation is that it is only necessary to
determine which side of the ``separatrix'' $\varphi=\pi$ the field
$\varphi(t)$ is on at the two (Euclidean) time values $t=0$
and $t=\beta/2$.  When we try to apply these ideas to the sphaleron
rate of QCD, we don't know how to measure the distance from the
separatrix, but it is perfectly feasible to determine which side of
the separatrix one is on.

One might worry that, in higher-dimensional problems like quantum
field theory, the separatrix is generally a codimension-1 surface,
rather than a single point.  Does this approach still work?  We argue
that it does.  At any point on the separatrix, one can find a local
orthogonal direction and ask about motion, and fluctuations, in that
direction.  At each such point, the speed a real-time configuration
moves through the separatrix is determined by the same kinetic
Hamiltonian as the Euclidean fluctuations orthogonal to the
separatrix. Therefore, their
relative amplitudes are proportional.  The real-time Affleck-type
calculation involves an integral over the separatrix surface of a
local probability density times the local
$\langle|\dd\varphi/\dd t|\rangle$, while our calculation involves the same
surface average of $\langle\sqrt{\Delta c^2}\rangle$, which is
proportional with fixed proportionality constant $1/2T$.
Therefore, to the extent that configurations close to the separatrix
really control the transition rate, the two calculations will be
equivalent when the factor $1/2T$ is included.

\section{Application to QCD}
\label{sec:QCD}

Let us see how to apply these ideas to the sphaleron rate in QCD.
First, we will consider one more time an evaluation in the toy model:

\subsection{$\Gamma$ in the toy model}
\label{sec:Gammatoy}

Before turning to QCD, suppose
we were trying to determine $\Gamma$ in the toy model,
and we had a nonperturbative method for studying the
finite-temperature Euclidean path integral.
However, suppose we somehow could not access the exact value of
$\varphi$, but only which side of $\pi$ it lies on.
We could perform our study as follows:
\begin{enumerate}
\item
  Generate a sample of typical Euclidean configurations:
\item
  For each, determine whether $\varphi$ is on the opposite side of
  $\varphi=\pi$ at $t=0$ as it is at $t=\beta/2$.
  This can be done by evolving $\varphi(t)$ under gradient-flow or
  relaxation dynamics and seeing whether it rolls down to $\varphi=0$
  or to $\varphi=2\pi$, for instance.
\item
  Evaluate the sphaleron rate as $2T$ times the fraction of
  configurations which make such a transition.
\end{enumerate}

\begin{figure}[htb]
  \includegraphics[width=0.45\textwidth]{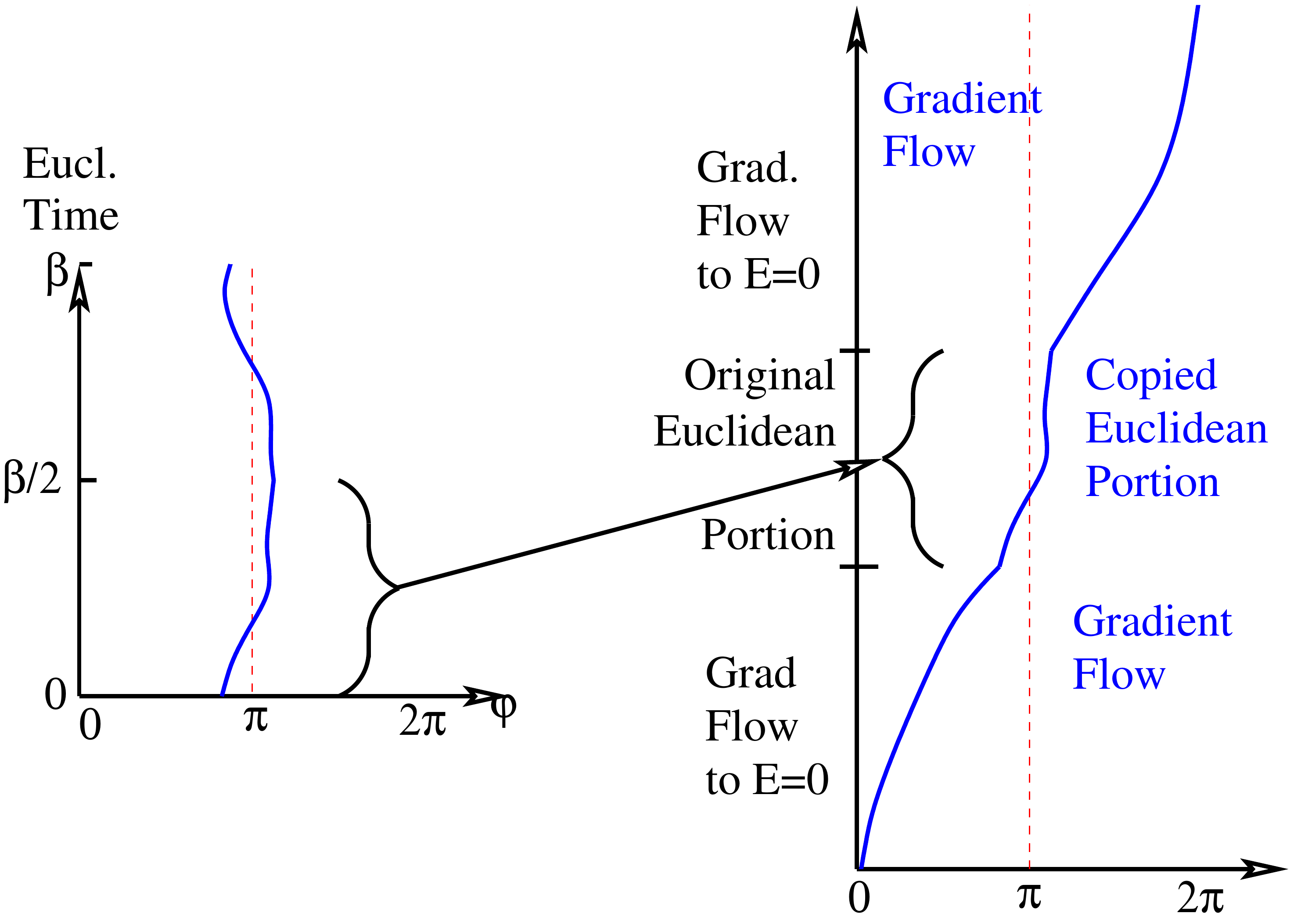}
  \hfill
  \includegraphics[width=0.45\textwidth]{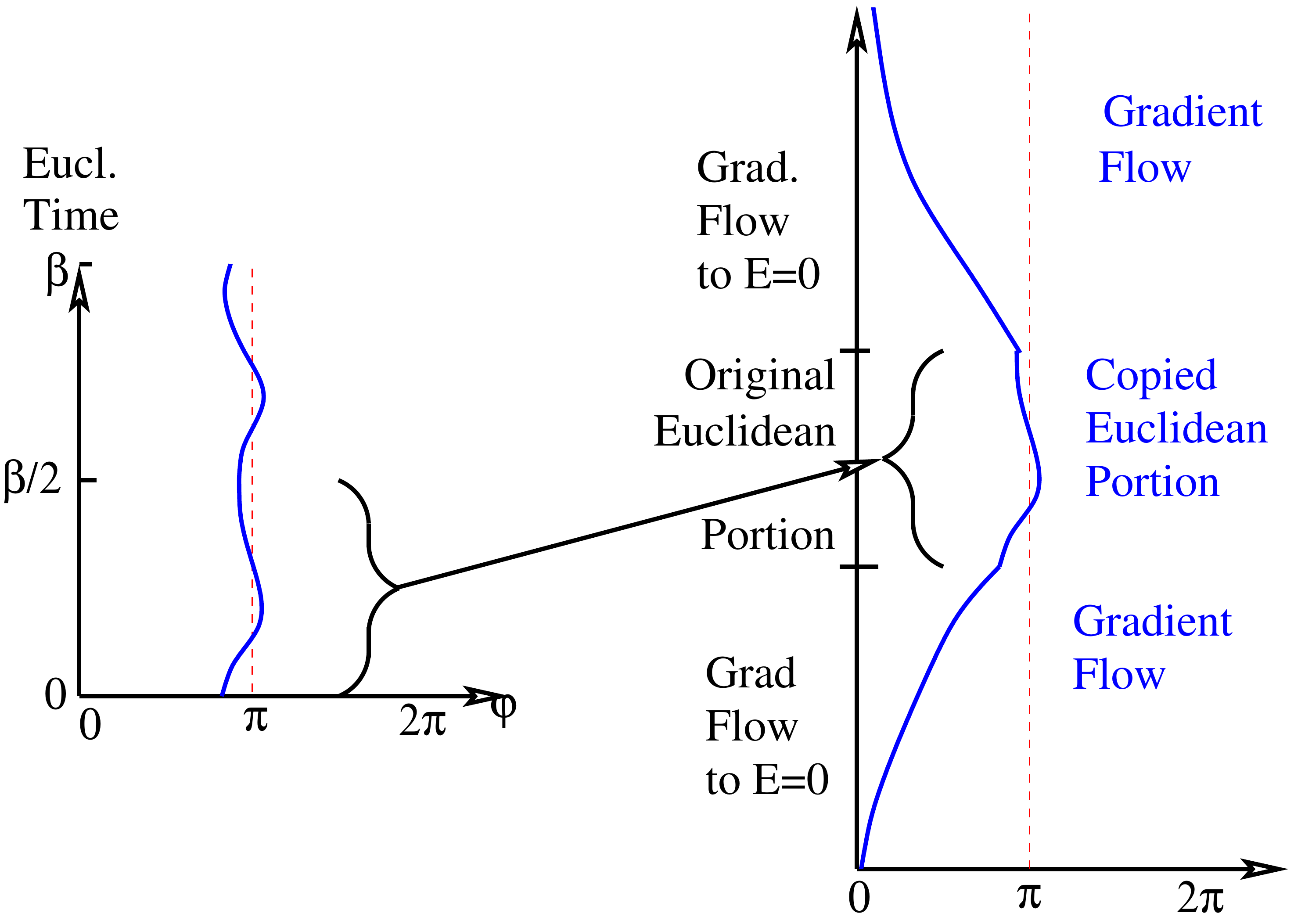}
  \caption{Cartoon of how two configurations would be evaluated using
    the gradient-flow approach described in the text.  A segment of a
    Euclidean simulation, from $t=0$ to $t=\beta/2$, is cut out of the
    simulation, and the starting/ending configurations are
    gradient-flowed to a zero-energy state.  The whole path
    (zero-energy to $t=0$ to $t=\beta/2$ to zero-energy) either has
    $\varphi$ change by $\pm 2\pi$ (left) or by 0 (right);
    in the former case it counts as a sphaleron. \label{fig:toyflow}}
\end{figure}

To understand the approach, consider Figure \ref{fig:toyflow},
which shows examples of how two configurations would be evaluated in
this procedure.

The section from $t=0$ to $t=\beta/2$ is cut from a simulation, and
one ``grafts'' the gradient-flow path from the $t=0$ configuration to
the vacuum and from the $t=\beta/2$ configuration to the vacuum onto
the two ends.  The full configuration starts and ends at the
zero-energy solution, $\varphi=0$,
\textsl{modulo} $2\pi$.  When the change is by $\pm 2\pi$, it is a
sphaleron; otherwise it is not.

\subsection{QCD: Chern-Simons number}
\label{sec:NCS}

Our approach in QCD will be the same:  to look for configurations
which cross a separatrix between $t=0$ and $t=\beta/2$, and therefore
settle into distinct vacua when acted on by gradient flow.
There is no easily-measured value $\varphi$ with a separatrix at
$\varphi=\pi$ in QCD.  However, we can use gradient flow and topology
to define a separatrix and to give a measure which determines whether
the $t=0$ slice and the $t=\beta/2$ slice are on opposite sides.

Let's start by defining gradient flow on a 3D configuration, in
analogy to its usual 4D definition \cite{Narayanan:2006rf,Luscher:2010iy}:
\begin{equation}
  \label{flowdef}
  B_i(t,\vec r , \tauft=0) = G_i(t,\vec r) \,, \qquad
  \frac{\partial}{\partial \tauft} B_i(t,\vec r , \tauft)
  = - \frac{\delta \int \dd^3 x \Tr G_{jk} G_{jk}[B]}{\delta B_i(\vec r)}
  \,.
\end{equation}
The fields $B_i(t,\vec r,\tauft)$, as a function of the 3D coordinate
$\vec r$ and the 3D flow-time coordinate $\tauft$, can be considered a 4D
gauge-field configuration in the temporal gauge.  One can then
integrate the topological charge density over this 4D configuration:
\begin{align}
  \label{Qflowdef}
  Q(t) & \equiv - \int_0^\infty \dd\tauft \int \dd^3\vec r \,\,
  q(t,\tauft, \vec r) \,, &
  q(t,\tauft,\vec r) & = \frac{\epsilon_{ijk}}{8\pi^2} \Tr G_{ij} \:
  \frac{\partial}{\partial \tauft} B_k(t,\vec r,\tauf) \,.
\end{align}
The result is equivalent to the definition of Chern-Simons number
adopted in Eq.~(5) of Ref.~\cite{Moore:1998swa},
 which predates papers discussing 4D gradient flow 
\cite{Narayanan:2006rf,Luscher:2010iy}.

To understand $Q(t)$ better, consider an idealized 4D instanton
solution centered at the origin, and consider a series of time slices
through the instanton.  For $t<0$ slices, gradient flow will make the
3D gauge field roll down towards the
vacuum described by the $t \to -\infty$ slice.  This will yield $Q(t)$
values which change from 0 at very negative $t$ towards $1/2$ for $t$
close to zero.  But on any $t>0$ slice, the gauge field
will gradient-flow down towards the vacuum described by the
$t \to +\infty$ slice.  Therefore, $Q(t)$ as defined above will have a
sharp jump at $t=0$ (the middle slice of the instanton), from a value
around $+1/2$ to a value around $-1/2$.
The slice where this jump occurs is our separatrix, and the property
of the separatrix is that such a 3D slice, under gradient flow, moves
to a saddlepoint between 3D vacua and sticks there.

\begin{figure}[htb]
    \centering
    \includegraphics[trim={0 1.5cm 0 2.2cm},clip,width=1\textwidth]{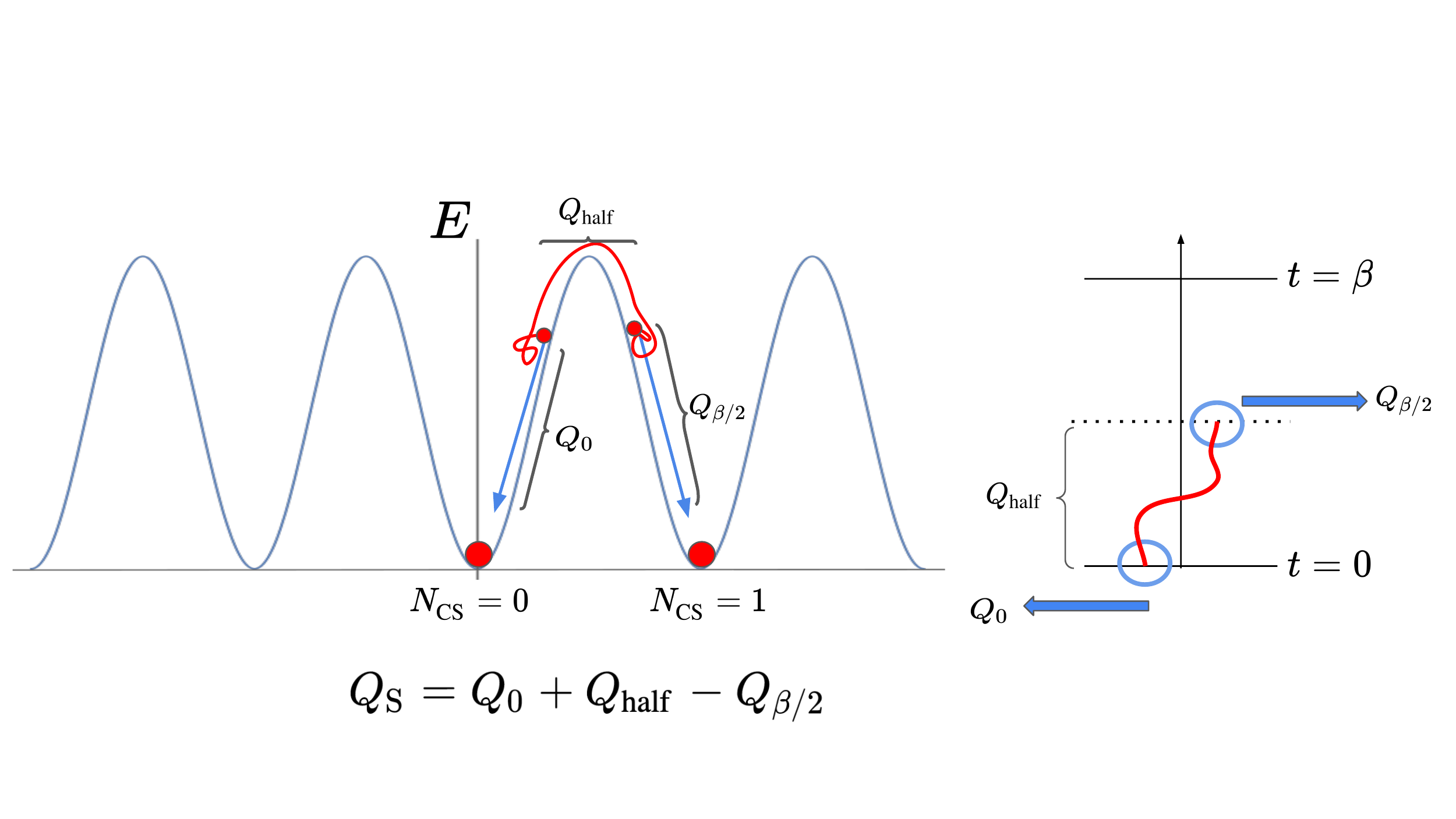}
    \caption{How we define and identify sphalerons in QCD.  The
      topological density is integrated between $t=0$ and $t=\beta/2$,
      and supplemented by its integration as these two boundaries are
      gradient-flowed to the vacuum.  When the configuration crosses a
      separatrix, the result is $\pm 1$.}
    \label{fig:toymodel_to_qcd}
\end{figure}

Now return to the periodic (thermal) Euclidean path integral of time
extent $\beta$.  We want to know whether the $t=0$ and $t=\beta/2$
slices are on opposite sides of this separatrix.  We could evaluate
$Q(t)$ as defined above at every time slice from $t=0$ to $t=\beta/2$
and see if it makes a sharp jump.  But this is inefficient.  If we
only want to know whether the $t=0$ and $t=\beta/2$ slices are on
opposite sides, we can consider the following combination:
\begin{align}
  \label{Qdef}
  \Qs & = Q(t=0) + \left[ \int_0^{\beta/2} \dd t \int \dd^3 x\: q(x,t)
 \right]  - Q(t=\beta/2)
  \\ \nonumber
  & \equiv Q_0 + \Qhalf - Q_{\beta/2} \,.
\end{align}
Here $\Qs$ stands for the ``sphaleron topology,'' not to be confused
with $\int_0^\beta \dd t \int \dd^3 x \: q(x,t) = \QI$ the instanton
topology of \Eq{NI}.
$\Qs$ is ``almost'' the difference between $Q(t)$, defined above, at
the two slices of interest.  If we cross the separatrix, we expect
$Q_0$ and $Q_{\beta/2}$ to be approximately $\pm 1/2$ and $\mp 1/2$ so
the difference is
$\pm 1$.  But by adding the $q$-integral over the intervening segment $\Qhalf$,
we make the measurement actually topological.  Namely, $\Qs$ is
the integral of $q(x,t)$ over a domain with no space boundaries (our
box has periodic boundary conditions), which begins on a vacuum
configuration at $t=0,\tauft=\infty$ and ends on a vacuum configuration
at $t=\beta/2,\tauft=\infty$, and has no field discontinuities in
between.  Therefore the integral of $q$, $\Qs$, is an integer.  When
this integer is nonzero, we define the configuration to be a sphaleron.
This idea is illustrated in Figure \ref{fig:toymodel_to_qcd}.

Lastly, note that the sphaleron rate is technically defined as the
number of topology transitions per unit \textsl{time and volume.}
Therefore, in defining the sphaleron rate, we must find the
mean-squared value of $\Qs$ from \Eq{Qdef}, multiply by $2T$, and divide
by the space volume which was studied -- that is,
$\langle \Qs^2 \rangle$ is expected to be extensive in the lattice
volume, provided the lattice is large enough to see the large-volume
behavior.

\section{Lattice calculation of the sphaleron rate}
\label{sec:strong-sph-rate}

As discussed in the last section, we will evaluate whether a given
lattice gauge-field configuration is a sphaleron by evaluating $\Qs$
defined in \Eq{Qdef}, determining
\begin{equation} \label{eq:lattice_connection}
    \frac{\GammaE}{T^3} = \frac{\expval{\Qs^2}}{\left(\Ns/\Ntau\right)^3}\,,
\end{equation}
with $\Ns$ and $\Ntau$ the number of lattice points in the space and
time directions respectively, and then rescaling by $2T$ to convert
this into the real-time rate as described in the appendix,
specifically \Eq{Convert}:
\begin{equation}
  \label{Convertcp}
  \Gsphal = 2T \GammaE \,.
\end{equation}

\begin{figure}[htb]
    \centering
    \includegraphics[trim={0 1.5cm 0 0.5cm},clip,width=1.0\textwidth]{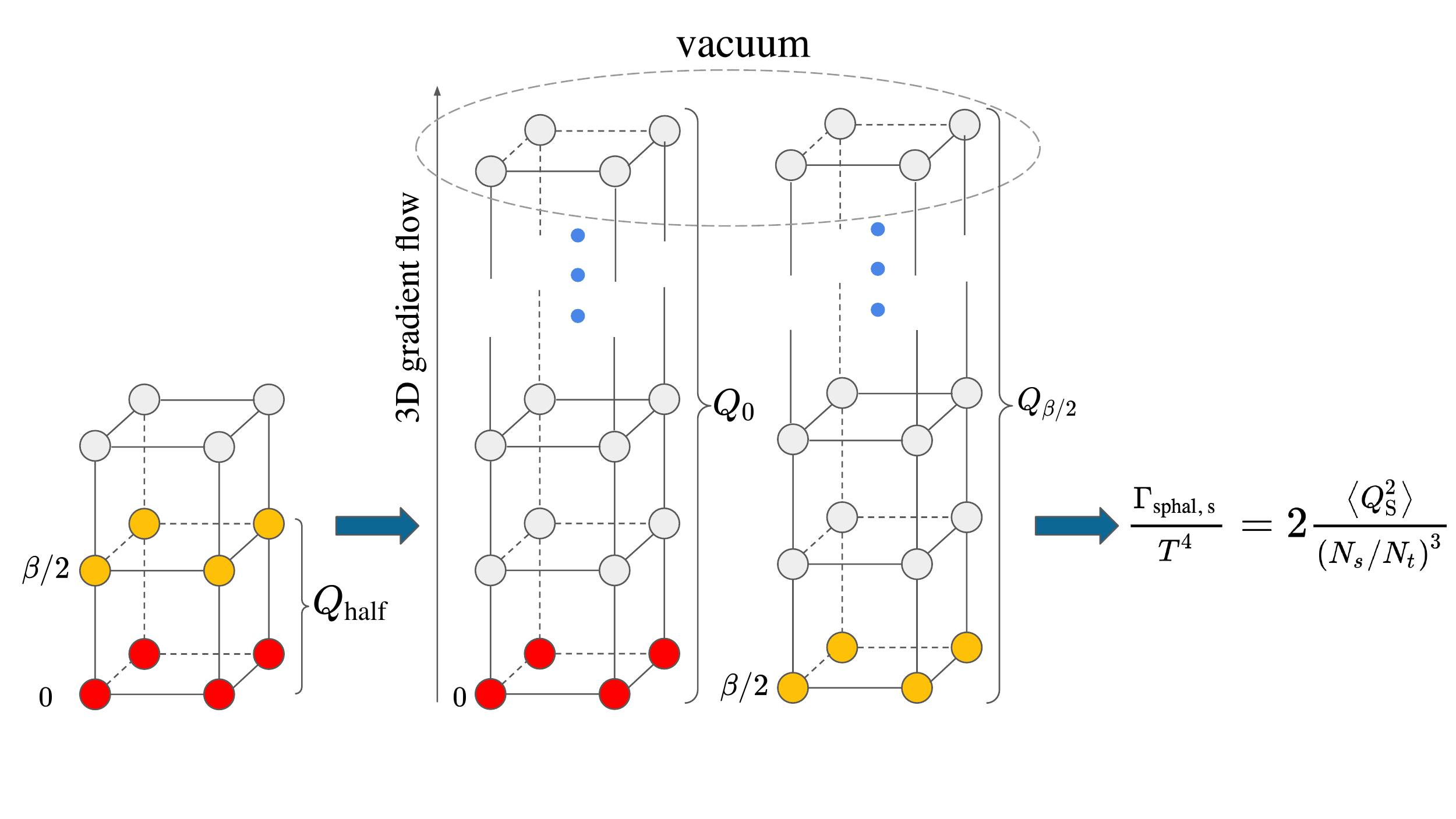}
    \caption{Sketch of the calculation performed on the lattice to determine the sphaleron rate. One dimension of the lattice is left not drawn for convenience.}
\label{fig:sketch_calculation}
\end{figure}

There is one additional complication which we must overcome to make
the procedure work.  Integrating $q(\vec r,t)$ on a lattice gauge
configuration typically returns a highly noisy result, because the
lattice definition of $q(\vec r,t)$ is contaminated with
high-dimension operators which are not topological.%
\footnote{We use the clover definition of $q(\vec r,t)$.}
Therefore it is essential to apply some lattice gradient flow (Wilson
flow \cite{Luscher:2010iy}) on the 4D configuration before the
evaluation.  This 4D gradient flow also reduces the size of the
fluctuations in the temporal direction which we discussed in
subsection \ref{sec:fluct} and in appendix \ref{sec:thm-euclidean}.
This will reduce how many configurations cross the separatrix and are
counted as sphalerons.  Appendix \ref{sec:toy-model-gf} contains a
calculation of the size of this effect, which we can use to correct
for the effects of gradient flow.

\begin{samepage}
Just to summarize the procedure, the steps that we need to take are
the following:
\begin{enumerate}
    \item Generate a valid lattice QCD configuration through a Hybrid
      Monte-Carlo algorithm.
    \item Apply a certain amount $\tauf$ of 4D gradient flow in order to
      reduce the UV noise.
    \item Calculate the topology enclosed in half the lattice.
    \item Extract the 3D slices at $\tE = 0$ and $\tE = \beta/2$, and
      compute $Q(0)$ and $Q(\beta/2)$ according to \ref{Qflowdef},
      using a 3D version of gradient flow all the way to the vacuum.
    \item Calculate $\Qs$ with these quantities, see
      \Eq{Qdef}. Repeat for many configurations to extract
      $\expval{\Qs^2}$.
\end{enumerate}
\noindent The procedure for determining $\Qs$ is illustrated in figure \ref{fig:sketch_calculation}.
\end{samepage}

In this paper, we have used openQCD \cite{openQCD} for generating the
configurations and applying the 4D gradient flow, and then implemented
the calculation of \ref{Qflowdef} separately.
We update with the HMC algorithm with trajectories of length 1.5 (in lattice 
units), measuring every 60 updates. The autocorrelation between
configurations is then acceptable for the $\Ns$ values we use.
For the statistical analysis we have used the $\Gamma$-method algorithm as
provided in \cite{DEPALMA2018} as a Python module.

When integrating the 3D slices along the flow direction, we have also
used blocking-techniques (as were described in \cite{Moore:1998swa})
after the energy had been lowered below a threshold (chosen such that
the blocking did not change the final answer).
It is also important to note that while the integration of the 4D
gradient flow discretization needs to be done as precisely as possible
(and for that, we have used the 3rd order Runge-Kutta algorithm already implemented in openQCD), the
integration for the 3D slices can be done with a much simpler Euler
algorithm with bigger step-size, as the most important part is that we
end up in vacuum (where all links are close to unity), and not the
precise configuration we pass through in the middle.

For the scale setting, we have used the fitting parameters found in
\cite{Francis:2015lha, Burnier:2017bod} which allows us to find the
$\betalatt$ necessary to equilibrate our lattice to our desired
temperature, given the number of sites in the time direction $\Ntau$.
For reference, we have included in table \ref{tab:blatt} the value 
of the lattice coupling for different cases we have used in this work.

\begin{table}[]
\centering
\begin{tabular}{cc|cccccc|}
\cline{3-8}
\multicolumn{1}{l}{}                           & \multicolumn{1}{l|}{} & \multicolumn{6}{c|}{$\Ntau$}                                                                                                          \\ \cline{3-8}
\multicolumn{1}{l}{}                           & \multicolumn{1}{l|}{} & \multicolumn{1}{c|}{6}    & \multicolumn{1}{c|}{8} & \multicolumn{1}{c|}{10} & \multicolumn{1}{c|}{12} & \multicolumn{1}{c|}{14} & 16 \\ \hline
\multicolumn{1}{|c|}{\multirow{8}{*}{$T/\Tc$}} & \multicolumn{1}{c|}{1.3} & \multicolumn{1}{c|}{6.04979} & \multicolumn{1}{c|}{6.23700} & \multicolumn{1}{c|}{6.39495} & \multicolumn{1}{c|}{6.53054} & \multicolumn{1}{c|}{6.64889} & \multicolumn{1}{c|}{6.75371}   \\ \cline{2-8}
\multicolumn{1}{|c|}{}                         & \multicolumn{1}{c|}{2}   & \multicolumn{1}{c|}{6.33718} & \multicolumn{1}{c|}{6.54976} & \multicolumn{1}{c|}{6.72273} & \multicolumn{1}{c|}{6.86803} & \multicolumn{1}{c|}{6.99311} & \multicolumn{1}{c|}{7.10283}   \\ \cline{2-8}
\multicolumn{1}{|c|}{}                         & \multicolumn{1}{c|}{3}   & \multicolumn{1}{c|}{6.64031} & \multicolumn{1}{c|}{6.86803} & \multicolumn{1}{c|}{7.04966} & \multicolumn{1}{c|}{7.20049} & \multicolumn{1}{c|}{7.32939} & \multicolumn{1}{c|}{7.44187}   \\ \cline{2-8}
\multicolumn{1}{|c|}{}                         & \multicolumn{1}{c|}{4}   & \multicolumn{1}{c|}{6.86803} & \multicolumn{1}{c|}{7.10283} & \multicolumn{1}{c|}{7.28847} & \multicolumn{1}{c|}{7.44187} & \multicolumn{1}{c|}{7.57254} & \multicolumn{1}{c|}{7.68632}   \\ \cline{2-8}
\multicolumn{1}{|c|}{}                         & \multicolumn{1}{c|}{5}   & \multicolumn{1}{c|}{7.04966} & \multicolumn{1}{c|}{7.28847} & \multicolumn{1}{c|}{7.47640} & \multicolumn{1}{c|}{7.63126} & \multicolumn{1}{c|}{7.76294} & \multicolumn{1}{c|}{7.87746}   \\ \cline{2-8}
\multicolumn{1}{|c|}{}                         & \multicolumn{1}{c|}{7}   & \multicolumn{1}{c|}{7.32939} & \multicolumn{1}{c|}{7.57254} & \multicolumn{1}{c|}{7.76294} & \multicolumn{1}{c|}{7.91939} & \multicolumn{1}{c|}{8.05216} & \multicolumn{1}{c|}{8.16748}   \\ \cline{2-8}
\multicolumn{1}{|c|}{}                         & \multicolumn{1}{c|}{10}  & \multicolumn{1}{c|}{7.63126} & \multicolumn{1}{c|}{7.87746} & \multicolumn{1}{c|}{8.06960} & \multicolumn{1}{c|}{8.22716} & \multicolumn{1}{c|}{8.36070} & \multicolumn{1}{c|}{8.47657}   \\ \cline{2-8}
\multicolumn{1}{|c|}{}                         & \multicolumn{1}{c|}{20}  & \multicolumn{1}{c|}{8.22716} & \multicolumn{1}{c|}{8.47657} & \multicolumn{1}{c|}{8.67054} & \multicolumn{1}{c|}{8.82926} & \multicolumn{1}{c|}{8.96359} & \multicolumn{1}{c|}{9.08003}   \\ \hline
\end{tabular}
\caption{\label{tab:blatt} Value of the lattice coupling $\betalatt$ for different values of $T/\Tc$ (where $\Tc$ is the critical temperature of the pure-glue Yang-Mills theory, given by $\Tc = 287.4(70)$ MeV) and number of sites in the temporal direction $\Ntau$.}
\end{table}

\begin{figure}[tb]
    \centering
    \includegraphics[trim={3cm 1.3cm 1cm 0.0cm},clip,width=1.05\textwidth]{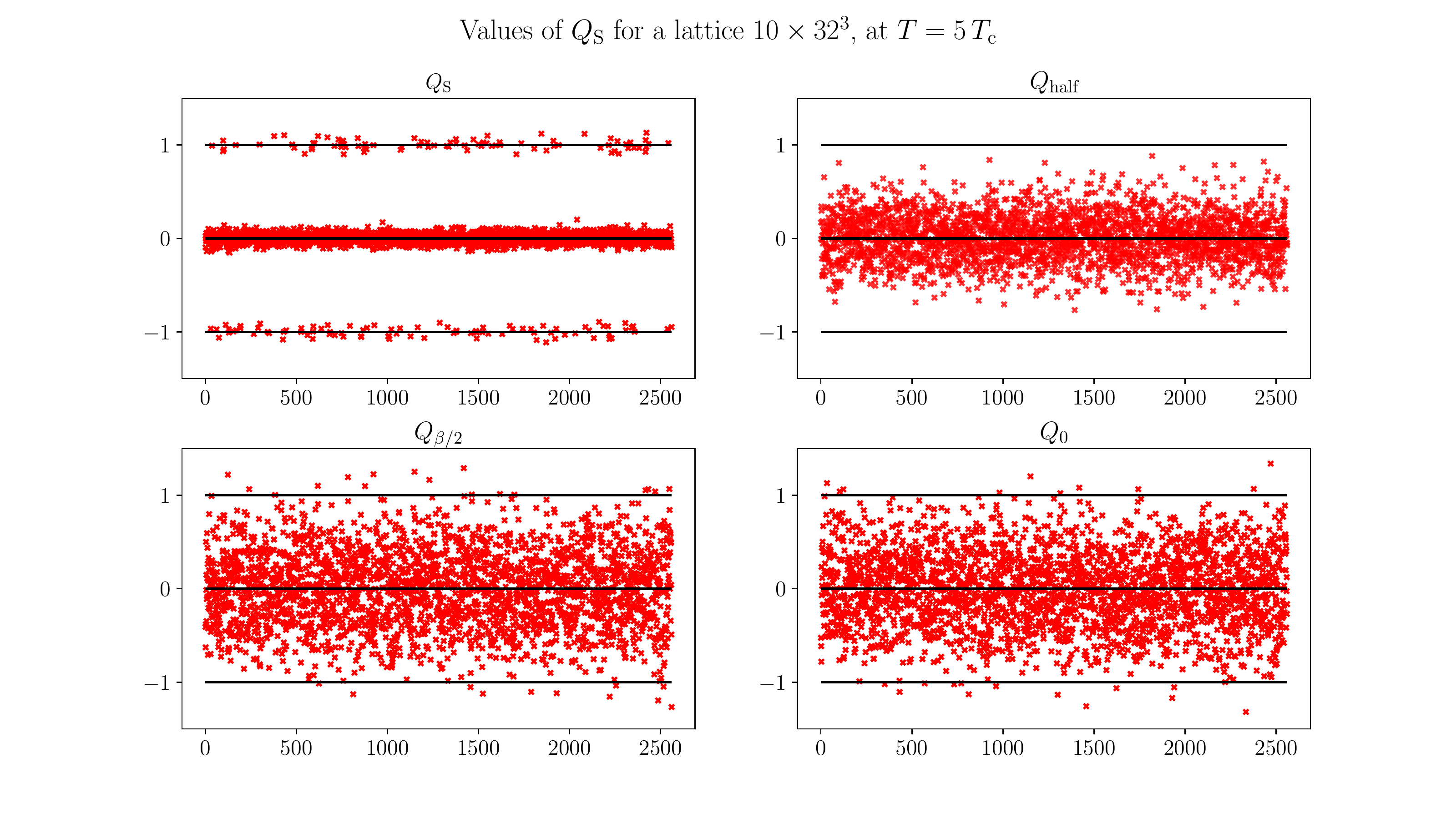}
    \caption{Numerical data obtained from 2560 configurations on a
      lattice with $10 \times 32^3$ sites, equilibrated at a
      temperature $T = 5 \,\Tc$, where the measurement has been
      performed at a gradient flow depth of $\tauf = 1.2 a^2$. The
      top-left plot is $\Qs$ of each configuration, and is
      approximately an integer. The top-right, bottom-left, and
      bottom-right plots show the $\Qhalf$, $Q_0$, and
      $Q_{\beta/2}$, respectively.}
    \label{fig:numerical_ncs}
\end{figure}

\begin{figure}[]
    \centering
    \includegraphics[trim={2cm 0.5cm 1cm 0.0cm},clip,width=1.05\textwidth]{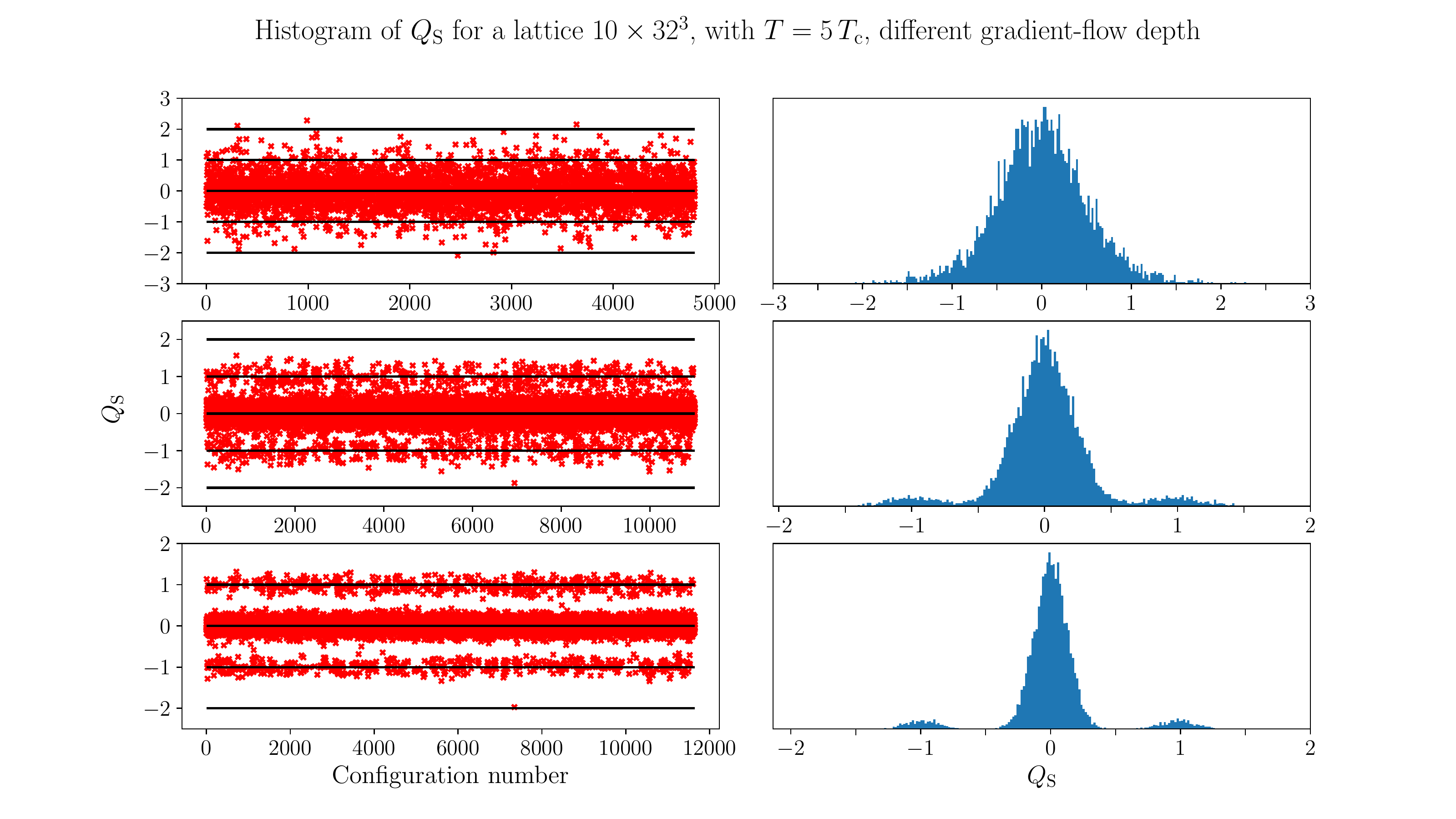}
    \caption{Histograms of $\Qs$ for a lattice with $10 \times 32^3$
      sites, at temperature $T = 5 \,\Tc$, and different gradient flow
      depths. The amount of flow corresponds to
      $\tauf=[0.2, 0.4, 0.6]\, a^2$,
      top to bottom. We see that as we go down the flow
      path, the peaks are more defined. While for $\tauf/a^2 = 0.2$ we
      cannot see the topological nature of the jumps, that becomes
      clear for $\tauf/a^2 = 0.4$. As a remark, the criterion
      \Eq{eq:criterion} fails for the top case, but is satisfied for
      the other two.}
    \label{fig:histograms_ncs}
\end{figure}

Before trying to understand the continuum limit of this method, and to
be sure we are on the right track, we first test that the procedure
really returns integer values of $\Qs$.  We do this on a single lattice
size and spacing and a single chosen value of 4D gradient flow depth
and show the results in \ref{fig:numerical_ncs}.  We see that, while
the individual components $Q_0, \, \Qhalf, \, Q_{\beta/2}$ are definitely not
topological and take a range of values, the combination $\Qs$ is indeed
topological up to modest fluctuations caused by $\OO(a^2)$
high-dimension operator contamination in our definition of topological
density.  This effect gets smaller as we apply more 4D gradient flow
(see below).  In practice we must apply enough 4D gradient flow to
cleanly separate the different integer values, which can be checked by
making a histogram of $\Qs$ values and seeing that they separate into
clear peaks.  Then one assigns integer values by projecting to the
closest integer.
We illustrate how this works in figure
\ref{fig:histograms_ncs}, where we see the histogram of $\Qs$ at
different gradient flow depths, and precisely check that the peaks get
sharper as we go down the flow path.

The quantitative criterion we use to determine whether enough 4D
gradient flow has been used is the following.  First, we consider all
configurations with $|\Qs| < 0.5$ and we evaluate the standard
deviation $\sigma^2_\text{data} = \langle \Qs^2 \rangle_{|\Qs|<0.5}$.
Assuming that the true distribution is Gaussian, we ask how much of
the distribution is expected to lie outside the range
$|\Qs|<0.5$, and we require that this be less than 2\%:
\begin{equation}
  \label{eq:criterion}
    \int_{1/2}^{\infty} \frac{2 \dd{z} }{\sqrt{2 \pi
        \sigma^2_\text{data}}} e^{- \frac{z^2}{2
        \sigma^2_\text{data}}} = \text{erfc} \left(\frac{1}{2 \sqrt{2}
      \sigma_\text{data}} \right) < 0.02
    \quad \to \quad \sigma_\text{data} <
    0.215 \,.
\end{equation}

\section{Volume, lattice spacing, and gradient-flow depth dependence}
\label{sec:tests}

Now that we have formulated the theoretical framework of the calculation and shown that it works numerically, we need to analyze how it behaves on the lattice.
The three main topics we should study are:
\begin{enumerate}
\item
  how to correct the rate for the effects of 4D gradient flow,
\item
  how it changes when we vary the spatial volume of our lattices, and
\item
  how to take the continuum limit, that is, how it changes when we go
  to finer lattices (by varying the number of sites in the time
  direction, and varying $\betalatt$ such that the temperature is
  constant).
\end{enumerate}

\subsection{Gradient flow dependence}
\label{sec:flowtest}

As explained in the previous section, some 4D gradient flow is needed to
ensure that $\Qs$ is really topological up to fluctuations which are
small enough to fix by projecting to the nearest integer.
However, applying this gradient flow can move individual time-slices
across the separatrix and reduces how many configurations are
determined to be sphalerons.  We have to correct for this effect.
The analytical dependence in the case of a single degree of freedom has already
been explored in section \ref{sec:toy-model-gf}, and here we will
check that the same relationship holds for the case of QCD on the
lattice.
By taking the quotient between equations \Eq{eq:sph_rate_result} and
\Eq{eq:sph_rate_flow}, we can find that
\begin{equation}
    \label{eq:gradient-flow-correction}
    \frac{\Gsphal^\tauf}{\Gsphal} = \frac{2}{\pi} \sqrt{2 \sum_{\substack{n=1 \\ n \text{ odd}}}^\infty \frac{e^{-\taufbar n^2}}{n^2}}
\end{equation}
where we have defined $\taufbar \equiv 8 \pi^2 \tauf/ \left( a^2 \Ntau^2 \right)$.
\begin{figure}[tbh]
    \centering
    \includegraphics[trim={1.8cm 0.0cm 2cm 0cm},clip,width=1.05\textwidth]{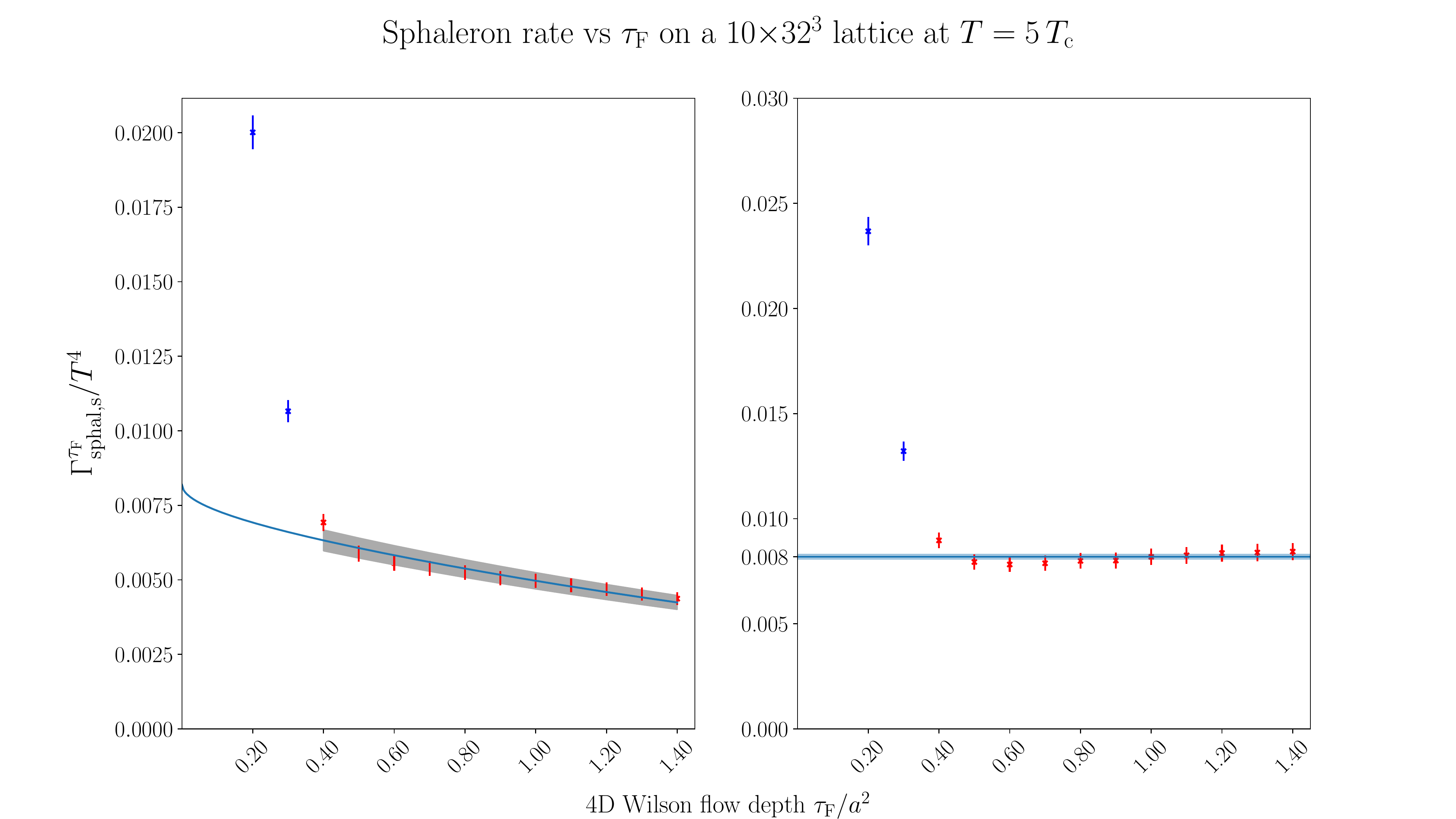}
    \caption{Left: measurement of the sphaleron rate at different
      gradient-flow depths, where the points in red (representing
      different values of $\Gsphal^\tauf$) have been used to find the best
      fit for $\Gsphal$ (see equation
      \Eq{eq:gradient-flow-correction}). The points in blue do not
      satisfy the criterion explained in a previous section, equation
      \Eq{eq:criterion}. Right: same plot but where each value has
      been corrected according to \Eq{eq:gradient-flow-correction},
      and the horizontal line is the best fit described before. We see
      that the correction successfully accounts for the effects of 4D
      gradient flow.}
    \label{fig:gradient-flow-sweep}
\end{figure}

We can then explore the accuracy of this correction by examining the
same ensemble of configurations using different amounts of gradient
flow.  The result can be seen in figure \ref{fig:gradient-flow-sweep}, where
the points in the left represent the sphaleron rate $\Gsphal^\tauf$,
while on the right they have been corrected according to
\Eq{eq:gradient-flow-correction}.  We see that, for gradient flow
depths above a threshold of about $\tauf = 0.5a^2$, the correction
accurately describes the effect of gradient flow.

We expect the correction to break down when $\tauf$ comes of order
$\beta^2$, that is, $a^2 \Ntau^2$, when the gradient flow starts
erasing all information stored in the gauge fields.
We have not explored this regime
in detail because we don't need to -- as long as there is a range of
$\tauf$ values where the correction works well, then we can use this
to establish the sphaleron rate in a $\tauf$-independent way.
This should be possible whenever $\Ntau$ is sufficiently large, which
is the same as a requirement that the lattice spacing be sufficiently
small.

\subsection{Volume dependence}
\label{sec:Voltest}

The sphaleron rate has been computed at very high temperatures (or
weak gauge couplings) using classical-field, real-time methods, and it
was always found that the rate per unit 4-volume is suppressed in
small boxes but grows to an infinite-volume asymptotic value above
some box size \cite{Ambjorn:1995xm,Moore:1999fs,Moore:2010jd}.
Intuitively, this means that sphalerons have a typical intrinsic size,
and the box must be larger than this size for a sphaleron to ``fit.''
At weak coupling, the size is expected to be parametrically of order
$1/g^2 T$ \cite{Kuzmin:1985mm,Arnold:1987mh}.
In the 4D Euclidean calculation we expect something similar to happen:
in lattices with $\Ntau$ sites in the temporal direction and $\Ns$
sites in the spatial direction, for small $\Ns$ we will not observe
any jumps.  As we make the spatial volume bigger (keeping $\betalatt$
and $\Ntau$ fixed), we will start observing more and more jumps.
Eventually, we would expect $\langle \Qs^2 \rangle / V$ to approach a
constant.

\begin{figure}[]
    \centering
    \hspace*{-1cm}
    \includegraphics[trim={1cm 0cm 3cm 0.5cm},clip,width=1.0\textwidth]{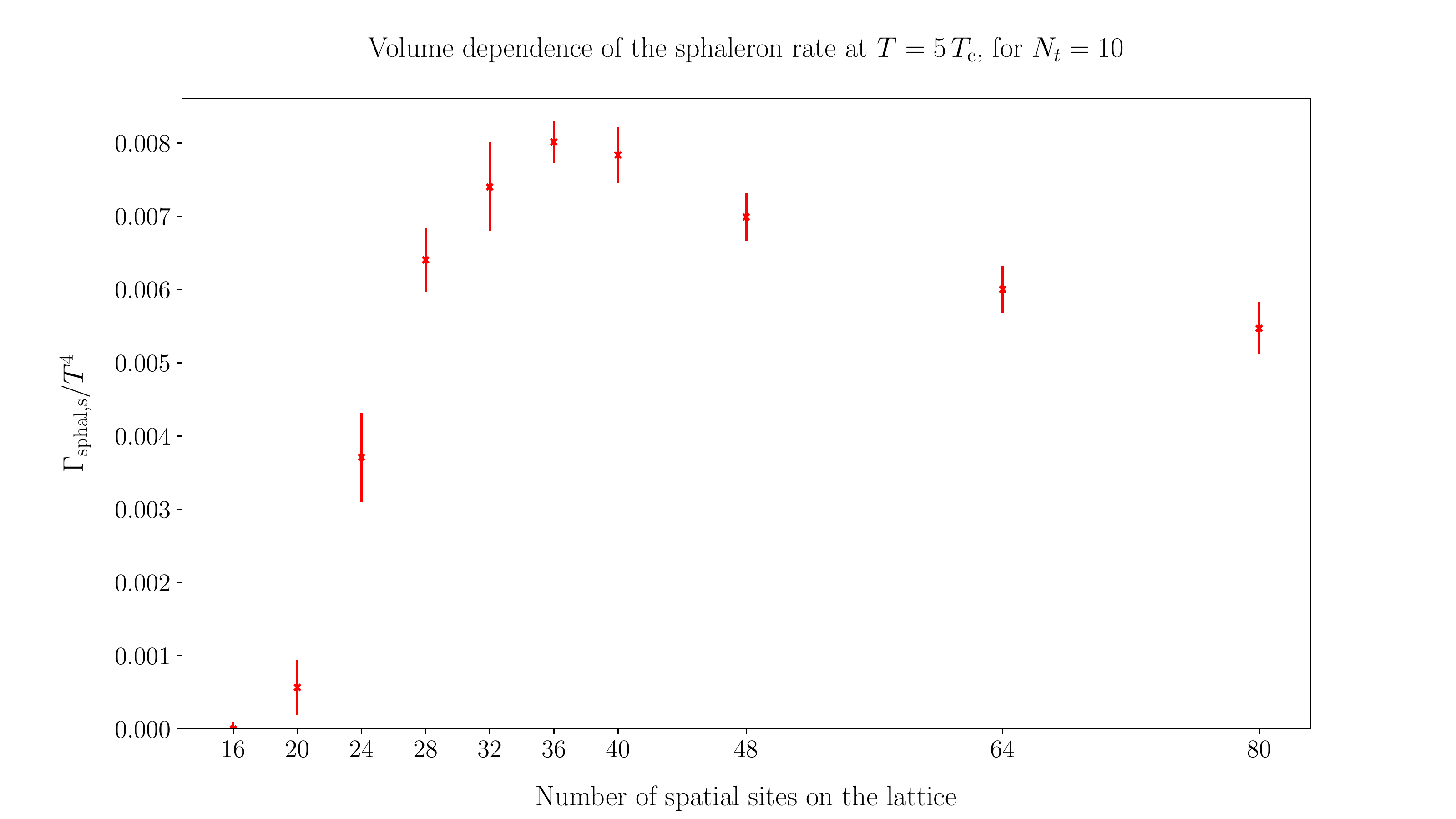}
    \caption{Sphaleron rate as a function of the aspect ratio
      $\Ns/\Ntau$, at fixed $\Ntau$ and temperature.}
    \label{fig:sweep_Ns}
\end{figure}

\begin{figure}[bt]
  \centering
  \includegraphics[trim={1cm 0cm 2.5cm 0.5cm},clip,width=1.0\textwidth]{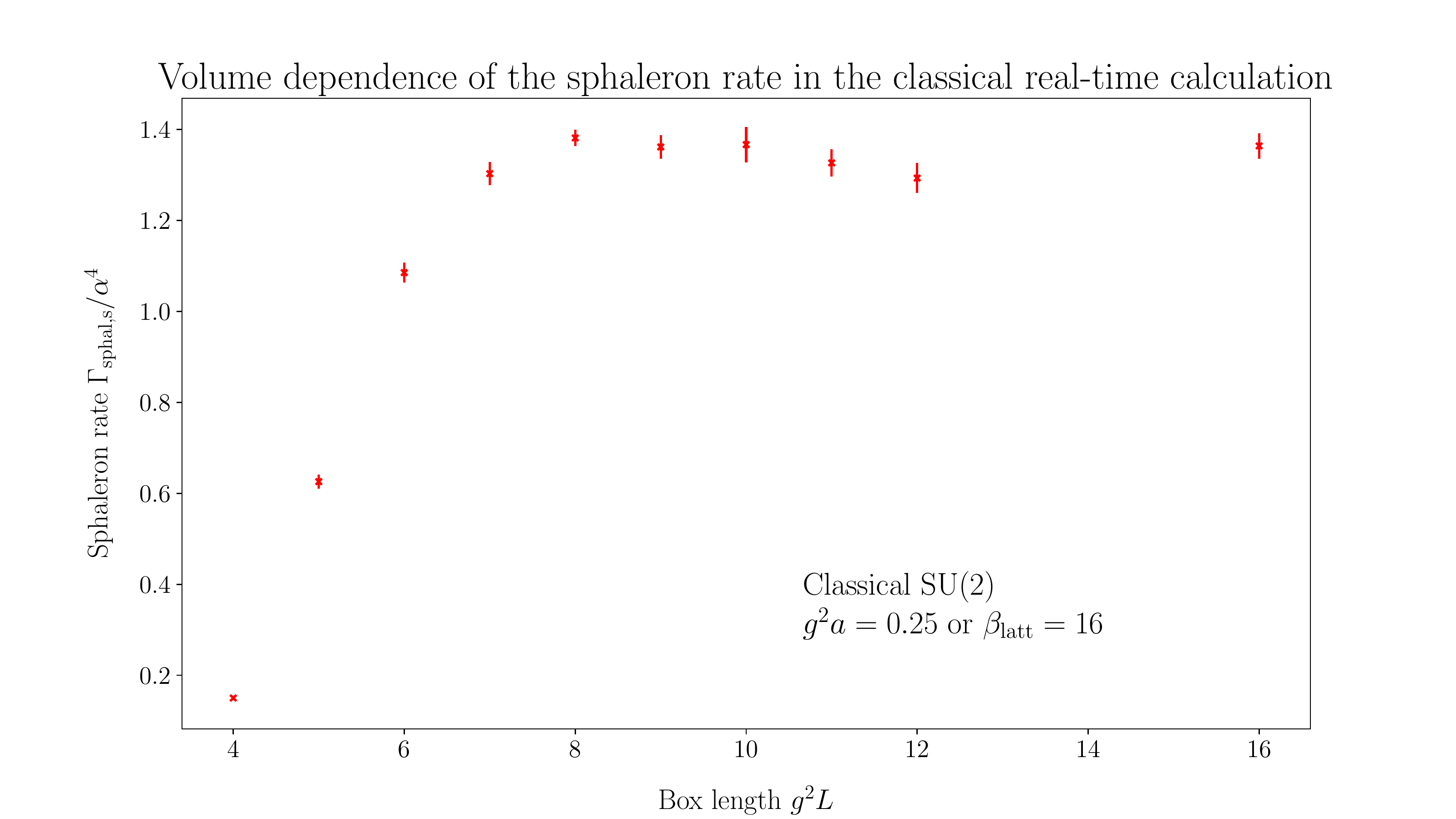}
  \caption{Volume dependence of the $SU(2)$ sphaleron rate when the
    rate is determined using classical-field, real-time techniques as
    in Ref.~\cite{Moore:1999fs}.  As the volume increases, the rate
    rises and saturates to a very clear plateau.}
  \label{fig:boxsize}
\end{figure}

We can see the result for different values of $\Ns$ in figure
\ref{fig:sweep_Ns}, where we have used a lattice with $\Ntau = 10$ and
equilibrated it at $T = 10 \,\Tc$.
As expected, we see the increase in the sphaleron rate as we increase
the volume until it reaches the peak.
But from that point forward, we see a decrease in the rate.
We did not expect this and it is not analogous to what occurs in the
real-time case.  To show this more clearly, we have revisited the
volume dependence of the classical real-time $SU(2)$ rate using the same
code as in Ref.~\cite{Moore:1999fs}.  Fixing the lattice spacing to
$a=0.25g^2 T$ and varying the box size, figure \ref{fig:boxsize} shows
that the classical real-time rate saturates to a flat asymptotic value
to within a few percent, without the clear post-peak decrease we
observe in the Euclidean method.

We do not yet understand why our sphaleron-rate volume dependence
exhibits a peak and a decrease towards very large volumes.  Note that
the configurations at larger volumes than the peak frequently exhibit
$|\Qs| > 1$, which means that there are multiple sphalerons in a box or
that our single-saddlepoint picture has broken down.
Since it is unclear if our approach is really appropriate in this
regime, we will use the peak in the volume-dependence curve as our
sphaleron-rate estimate in the remainder of this work.  Clearly it
would be valuable to revisit this decision and to find a clearer
understanding of the large-volume behavior and the meaning of this
peak and falloff.  We will assume that our results suffer an
$\OO(30\%)$ systematic uncertainty because of this assumption.

\subsection{Continuum limit}
\label{sec:atest}

Now the only thing we need to check is how $\Gsphal$ approaches the
continuum limit $a \to 0$ at fixed temperature and aspect ratio.
We do this by performing the calculation at a series of lattice
spacings (or $\Ntau$ values) and seeing how they scale with $a^2$.
As has been explained in section \ref{sec:strong-sph-rate}, we use the
scale setting as described in \cite{Francis:2015lha, Burnier:2017bod}
in order to relate the lattice spacing $a$ to the lattice coupling
$\beta$.  Then we can fix the temperature while increasing $\Ntau$
and calculate the sphaleron rate at the peak in each case%
\footnote{We also checked that the aspect ratio $\Ns/\Ntau$ where the
rate reaches its peak is independent of lattice spacing.},
using the previously established method to correct for 4D
gradient-flow effects.

\begin{figure}[]
    \centering
    \hspace*{-1cm}
    \includegraphics[trim={1.2cm 0.5cm 2.5cm 0.2cm},clip,width=1.0\textwidth]{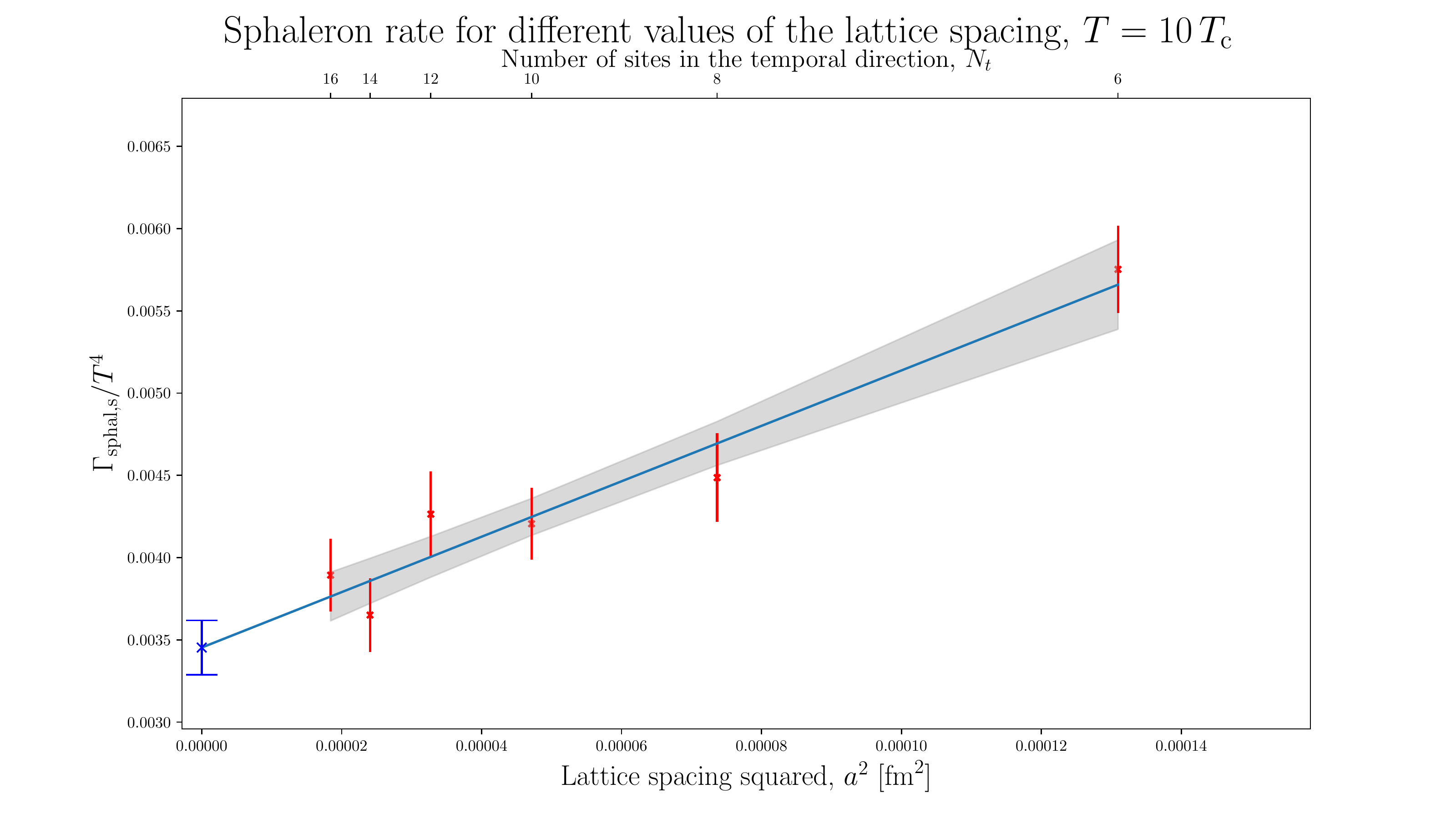}
    \caption{Sphaleron rate calculated at the peak of each
      volume-dependence curve, extrapolated to zero
      gradient-flow-depth, at different values of lattice spacing $a$,
      for $T$ = 10 $\Tc$. The zero-spacing extrapolation yields a
      value of $\Gsphal/T^4 = 0.0035 \pm 0.0002$.}
    \label{fig:continuum-limit}
\end{figure}

The result can be seen in figure \ref{fig:continuum-limit}.
As frequently happens, lattice spacing effects are large at $\Ntau=6$,
but for larger $\Ntau$ we find a good convergence towards a continuum
limit.

Rather than repeat this continuum limit procedure at every
temperature, we will fix to $\Ntau=10$ at all other temperatures,
which we see will introduce at most $\sim 20\%$ errors in our
results. 

\section{Low and high temperatures}
\label{sec:instanton}

Before we proceed to show the final results for the sphaleron rate at
different temperatures, we also need to comment on the limits of our
method.

\subsection{Low temperatures and instantons}
\label{subsec:instanton}

Consider the toy model one more time.  In the high-temperature regime,
the $\varphi$ field typically reaches $\varphi=\pi$ as a result of a
Euclidean history which loiters near $\varphi=\pi$ (a sphaleron),
rather than a history which transitions from $\varphi=0$ to
$\varphi=2\pi$ (an instanton).  But as the temperature is lowered, the
instantons become more important.  In this case, the basic picture --
that the real-time rate is controlled by classical thermal
fluctuations up to the barrier and the rate can be computed from the
frequency of such transitions over the barrier -- comes into
question.  Therefore we should not trust our approach if instantons
are as common as sphalerons.

One feature of our approach, as one sees in Figure \ref{fig:toyflow}
for instance, is that a sphaleron will give the same answer whether we
include the Euclidean path between $t=0$ and between $t=\beta/2$ as we
would find if we include the Euclidean path between $t=\beta/2$ and
$t=\beta$.  We see this more explicitly in \Eq{Qdef}.  If we had
chosen $t=\beta/2$ and $t=\beta$ as the lower and upper boundaries of
our region, we would define
\begin{align}
  \Qs' & = Q(t=\beta/2)
  + \left[ \int_{\beta/2}^\beta \dd t \int \dd^3 x \: q(x,t) \right]
   -  Q(t=\beta)
  \nonumber \\ & = Q_{\beta/2} + Q_{\mathrm{second-half}} - Q_0
  \nonumber \\ &
  = - ( Q_0 + Q_{\mathrm{half}}  - Q_{\beta/2} ) +
  \left[ \int_0^\beta \dd t \int \dd^3 x \: q(x,t) \right]
  \nonumber \\ &
  = - \Qs + \QI \,.
\end{align}
In passing from the first to second line we use periodicity, which
means that the configuration at $t=\beta$ is the same as at $t=0$ so
$Q_\beta = Q_0$.  The integral on the third line is the total topology
in the box, $\QI$.  At high temperatures $\QI$ is almost always zero,
and therefore $\Qs' = -\Qs$; the
topology we find in the upper half of the box is minus the topology we
find in the lower half.  But for configurations with an actual
instanton, $\QI = \pm 1 = \Qs + \Qs'$.
Therefore, true sphalerons ($\Qs=\pm 1$ but $\QI = 0$)
will count as a sphaleron
whether we use the top or the bottom half of our box, whereas
instantons will \textsl{either} count as a sphaleron if we use the top
half, \textsl{or} if we use the bottom half.

As we just emphasized, the whole approach of counting sphalerons and
equating them to classical transitions becomes suspect when a large
fraction of what we find are actually instantons.
Therefore it is useful to investigate the ratio of the instanton
density to the sphaleron rate.  When this ratio becomes of order
$1/2$, we have left the regime where classical activation makes
sense.  When the ratio is small, sphalerons dominate instantons and
our approach should make sense.

\begin{figure}[]
    \centering
    \hspace*{-1.8cm}
    \includegraphics[trim={2cm 0cm 3cm 0cm},clip,width=1.0\textwidth]{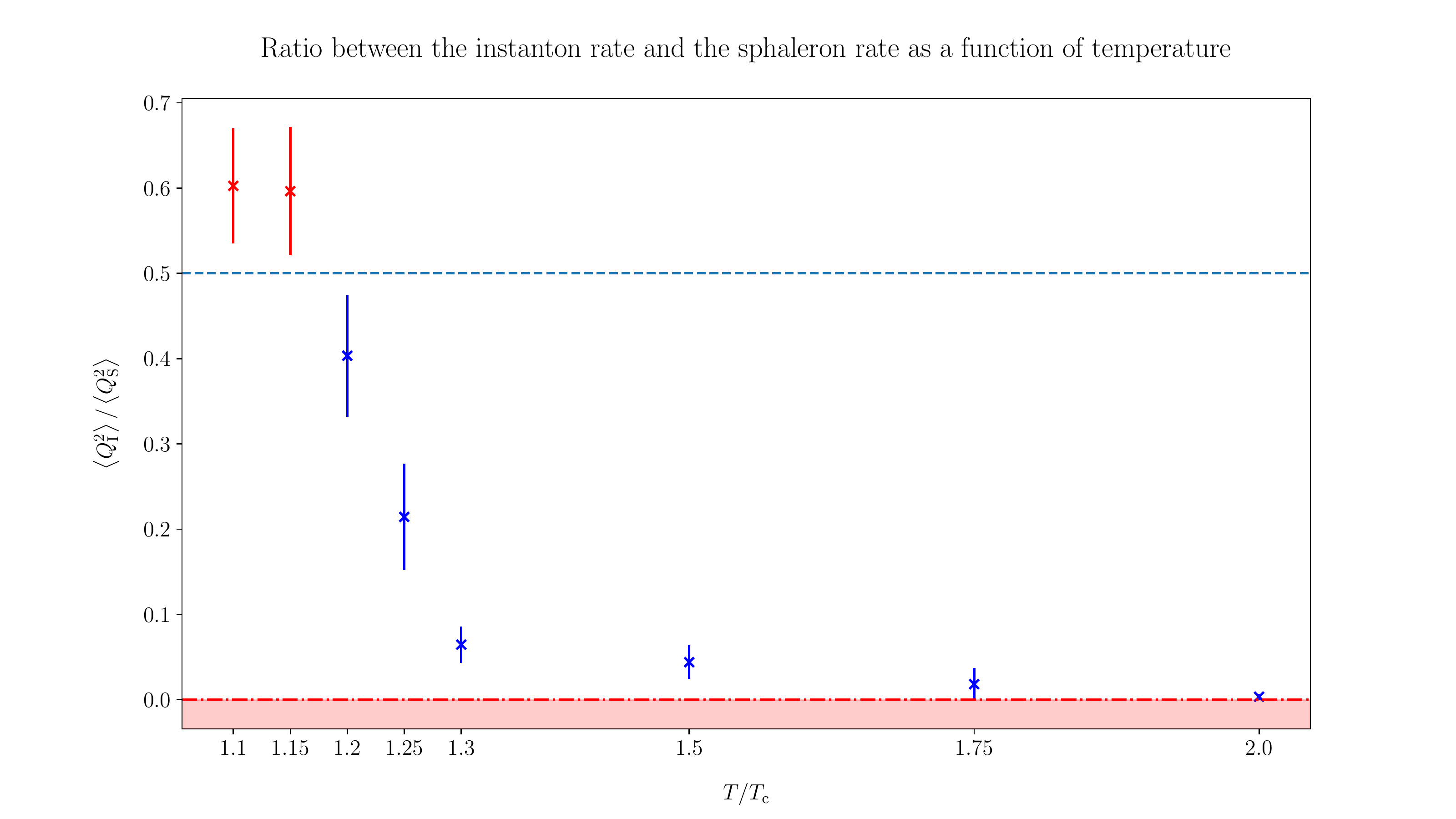}
    \caption{Ratio between the amount of instantons ($\QI \neq 0$) and
      sphalerons ($\Qs \neq 0$) as a
      function of temperature (in $\Tc$ units). Only the sphaleron
      rate has been extrapolated to zero gradient-flow-depth, as
      described in the previous section.}
    \label{fig:instantons_sphalerons}
\end{figure}

We present this comparison in Figure \ref{fig:instantons_sphalerons}.
For values of temperatures above 1.75 $\Tc$ the appearance of
instantons is exponentially suppressed and finding one is very rare.
For values between 1.3 and 1.75 $\Tc$, while we can find some
instantons, they are rare when compared to the amount of sphalerons
(sphalerons are at least 10 times more common than instantons).
For values lower than 1.3 $\Tc$, the instantons are very common and
therefore we cannot trust our calculation for the sphaleron rate.
Therefore, our method will only be reliable above 1.3 $\Tc$, and we
will restrict ourselves to this domain.

\subsection{High temperatures and classical fields}

\label{sec:comparetoclassical}

At very high temperatures the gauge coupling becomes so weak that
$g^2 T \ll T$.  In this case, sphalerons are physically large objects
which should be well described in terms of classical fields, both
thermodynamically and (within limits) dynamically.
In this regime, both our methods and previous, classical-field
real-time methods should be applicable.
Therefore we will attempt to make contact with the 3D real-time
results of Ref.~\cite{Moore:2010jd}.

Real-time results are calculated in terms of the effective 3D gauge
coupling $g_3^2$, which equals $g^2 T$ of the 4D theory at tree
level.  At the loop level, it can be related to the 4D $\MS$ coupling
via a perturbative matching calculation, which has been carried out to
2 loops by Laine and Schroeder in Ref.~\cite{Laine:2005ai}, see
Eq.~(2.34) to (2.37) of that reference.

The $\MS$ coupling can in turn can be related to the lattice coupling
$\betalatt$ by measuring the
gradient-flowed expectation value of the squared field strength
$E \equiv \Tr G_{\mu \nu} G^{\mu \nu}$ and using its NNLO relation to
the $\MS$ coupling at $\mu = 1/\sqrt{8\tauf}$ \cite{Harlander:2016vzb}:
\begin{equation}
  \label{measurecoupling}
    t^2 \expval{E(t)} = \frac{3 \alphas}{4\pi} \left(1 + k_1 \alphas + k_2 \alphas^2 \right)
\end{equation}
where $k_1 \approx 1.098$ and $k_2 \approx -0.982$ for SU(3) theory
without fermions.
We use the Wilson action, Zeuthen flow \cite{Ramos:2015baa}, and
clover definition of the field strength, checking that the result is
stable over a range of flow depths after the 2-loop running of
$g^2(\mu_{\MS})$ is taken into account.

Equipped with this method, we study a lattice with $\Ntau=8$ and
$\betalatt = 15$, finding that it corresponds to
$\frac{g_3^2}{4\pi} = 0.0406$.
Since the coupling is small, the box size must be correspondingly
large before the sphaleron rate ``turns on.''  For this choice of $\Ntau$, we find
that we need $\Ns = 96$ to reach the peak in the volume dependence of
the sphaleron rate.  Generating many independent configurations of
this size is challenging, so our statistical errors are rather large;
we find that $\Gsphal^{\text{new}}/(\alphas^4 T^4) = 19 \pm 7$.
Based on figure \ref{fig:continuum-limit}, we expect the real answer
to be $20\%$ to $25\%$ lower once the continuum limit is performed.
Going back to \cite{Moore:2010jd}, we find that for $\alphas \text{(0
  flavor)} = 0.039$, the value is
$\Gsphal^{\text{old}}/(\alphas^4 T^4) = 11.5 \pm 0.3$.
Therefore, our method is compatible with the previous 3D
calculations, given the downwards continuum corrections, the rather
large statistical errors, and the theoretical errors
present in the calculation in \cite{Moore:2010jd}.

\section{Results, Discussion, and Conclusions}
\label{sec:results}

\begin{figure}[]
    \centering
    \hspace*{-1.1cm}
    \includegraphics[trim={2.5cm 0cm 3.5cm 0cm},clip,width=1.1\textwidth]{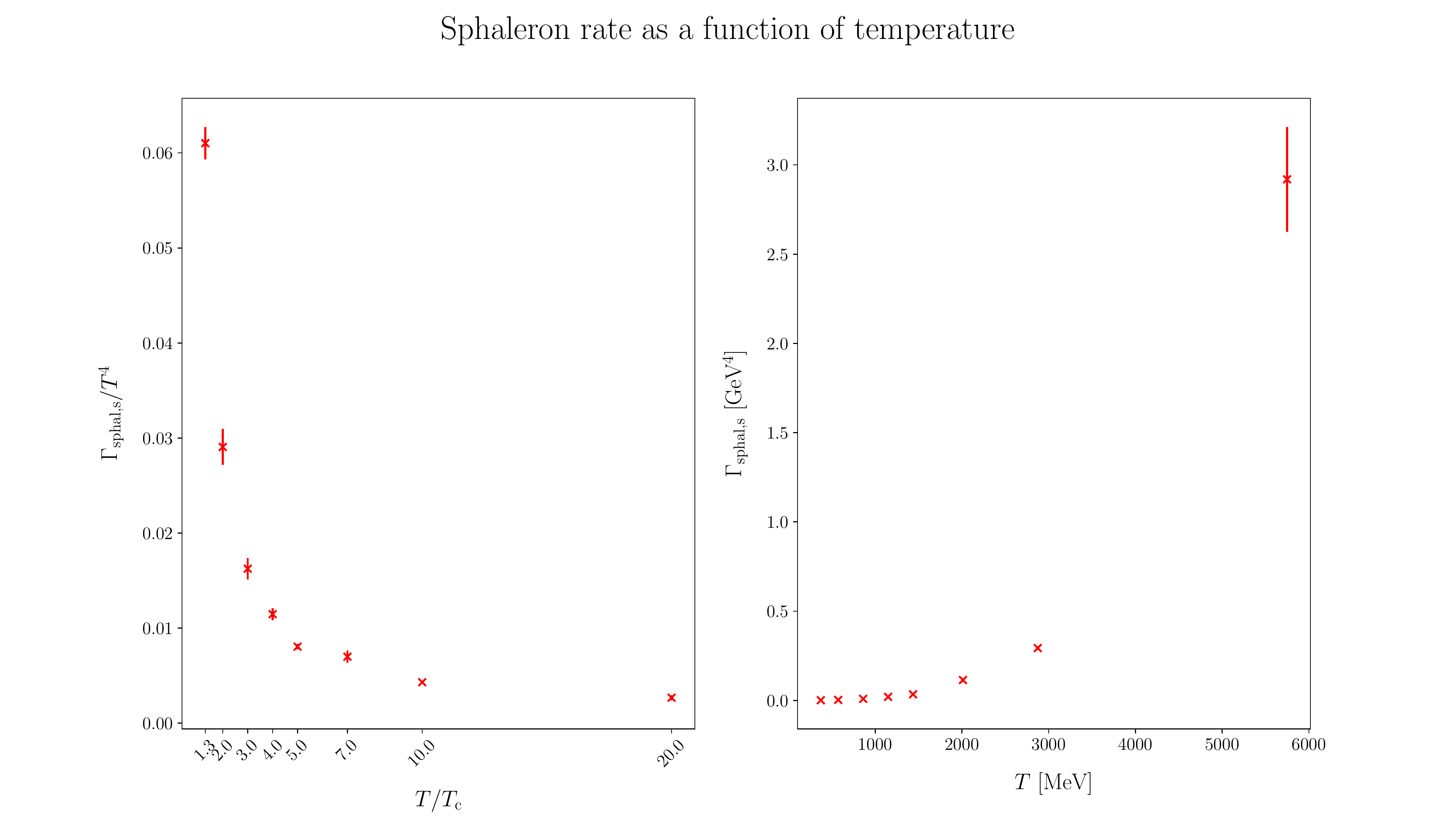}
    \caption{Results for the sphaleron rate for different
      temperatures.
      The left plot presents the dimensionless ratio
      found by scaling with $T^4$, while on the right we have
      reintroduced dimensions, using $\Tc \approx 280$ MeV for SU(3)
      theory.  Errors are statistical only; all data are at
      $\Ntau=10$ without a continuum extrapolation.}
    \label{fig:sph_rate}
\end{figure}

\begin{table}
  \centering
  \begin{tabular}{|c|c|c|c|}
    \hline
    $T/\Tc$ &    $\Gsphal/T^4$ &      $\Gsphal[\mathrm{GeV}^4]$ & Num. configs. \\ \hline
    1.3  &  $ 0.061   \pm 0.002 $  & $ 0.00119 \pm 0.00003 $ & 5520 \\
    2.0  &  $ 0.029   \pm 0.002 $  & $ 0.0031  \pm 0.0002  $ & 1970 \\
    3.0  &  $ 0.016   \pm 0.001 $  & $ 0.0089  \pm 0.0006  $ & 1280 \\
    4.0  &  $ 0.0115  \pm 0.0006$  & $ 0.020   \pm 0.001   $ & 2260 \\
    5.0  &  $ 0.0080  \pm 0.0003$  & $ 0.034   \pm 0.001   $ & 7680 \\
    7.0  &  $ 0.0069  \pm 0.0007$  & $ 0.11    \pm 0.01    $ & 1280 \\
    10.0 &  $ 0.0043  \pm 0.0002$  & $ 0.29    \pm 0.01    $ & 4560 \\
    20.0 &  $ 0.0027  \pm 0.0003$  & $ 2.9     \pm 0.3     $ & 1280 \\ \hline            
  \end{tabular}
  \caption{\label{tab:sph_rate} Results presented in figure
    \ref{fig:sph_rate} in tabular form.  Errors are statistical only.}
\end{table}

The results obtained with our method for the sphaleron rate in the
range of temperatures $[1.3 - 20]\,\Tc$ are shown in figure
\ref{fig:sph_rate} and summarized in table \ref{tab:sph_rate}.
We emphasize that these results are at $\Ntau = 10$; we have not taken
a continuum limit.

The main application of these rates is to use them in \Eq{dndt} to see
how quickly or slowly a chiral imbalance relaxes via thermal sphaleron
processes.  Using $\Gsphal/T^3 \simeq 0.04 T$ over the range
$[1.3\Tpc,2\Tpc]$ and treating the up and down flavors as light ($n_f = 2$),
we estimate that the exponential decay lifetime for axial number is of
order $12/T$.  This is slow enough to allow axial number to play a role
in early dynamics during heavy ion collisions, but marginally fast
enough for axial number to relax before the system hadronizes.
Note that this estimate involves using pure-glue results in full QCD,
and it therefore still has rather large theory errors.

We have also calculated the sphaleron rate at the electroweak scale, 
where the coupling is approximately $\alphas \sim 0.1$. For that we 
have used the same techniques as in section \ref{sec:comparetoclassical} 
to find the 4D coupling from gradient flow. For $\betalatt = 9$, 
we have found that it corresponds to $\alphas = 0.108 \pm 0.002$ (where the  
uncertainty shown is not statistical but comes from having to pick a flow depth such that
equation \ref{measurecoupling} is valid but lattice errors are small). 
In this case, we obtain 
\begin{equation}
    \Gsphal(\alphas \sim 0.108) = 14 \pm 1 \, \alphas^4 T^4.
\end{equation}
We emphasize again that the error is statistical only; we have not
taken a continuum limit and that the result is for pure glue, so there
are still much larger systematic effects to be included.

The main breakthrough of this paper is to show how real-time processes
mediated by semiclassical saddlepoints, like thermal bubble nucleation
or the sphaleron rate, can be calculated using exclusively Euclidean
and fully nonperturbative tools.  We have applied this technique to
the sphaleron rate in $SU(3)$ gauge theory partly because of its
phenomenological interest and partly because it is particularly clear
in this case how to define the separatrix and how to determine whether
the $t=0$ and $t=\beta/2$ time slices of the Euclidean path integral
lie on opposite sides of the separatrix.  The same approach should be
applicable to other saddle-controlled tunneling processes such as
bubble nucleation, though identifying the saddlepoint in this case may
be more challenging.
Our approach also has the weakness that it does not give us a way to
determine a dynamical prefactor $\ddd$ which appears in the tunneling
rate.  Note that this same defect is present in analytical Euclidean
approaches such as Affleck's method as well.  We currently don't see a
way to overcome this limitation.

It should be straightforward to extend the current approach to full
QCD with 2+1 light flavors of fermions.  The only change needed is to
include the fermions when performing the HMC lattice updates, so that
the configurations are drawn from the full, rather than pure-glue,
lattice ensemble.  We have already begun work on this project.

There are a few points where it would be useful to deepen our
understanding of the technique and our results.  The unexpected
scaling with lattice volume needs to be better understood.  And it
would also be interesting to better understand lower temperatures
where instantons start to play a role.  To what extent can we
incorporate instantons into a determination of axial number relaxation?
And to what extent does axial quark number make sense in a regime
where the system is moving towards a description in terms of hadronic
degrees of freedom?  We consider these to be interesting open issues.

\section*{Acknowledgments}
The authors acknowledge support by the Deutsche Forschungsgemeinschaft (DFG, German Research Foundation) through the CRC-TR 211 'Strong-interaction matter under extreme conditions'– project number 315477589 – TRR 211.
Calculations were performed on the high-performance computer Lichtenberg at the NHR Centers NHR4CES at TU Darmstadt.

\appendix

\section{Size of Euclidean fluctuations}
\label{sec:thm-euclidean}

We have argued that the sphaleron rate can be determined by finding
which Euclidean configurations ``span'' the separatrix, in the sense
that the spatial slice at $t=0$ and the slice at $t=\beta/2$ are on
different sides of the separatrix.

Here we present the details of the calculation comparing the
likelihood for a Euclidean configuration to satisfy this condition to
the real-time rate for transitions as determined by the Affleck
method.  We also incorporate the effect of 4D gradient flow applied to
the Euclidean configuration before the measurements are carried out.

The main assumption behind the method is that the sphaleron, critical
bubble, or other object responsible for tunneling is a robust
high-action object.  Even though the transition rate may not be
dominated by configurations close to the saddle point, it will be
dominated by configurations in the vicinity of the (codimension-1) separatrix
surface.  At every point on the separatrix there is a direction
orthogonal to the separatrix, and we assume that fluctuations along
this direction, when comparing 3D slices at different $t$ values, are
small relative to the overall size of the sphaleron.
In this case the degree of freedom $\phi(t)$ associated with this
direction can be treated as a single variable with a kinetic term
whose strength is approximately $\phi$-independent and a potential
which can be approximated as quadratic.  (We do \textsl{not} need this
kinetic term or this potential to be the same everywhere on the
separatrix -- we only need to know that there is little variation
orthogonal to the separatrix, over the field-space distances explored
by the higher Matsubara modes of this degree of freedom.)
If these assumptions are far from correct, the entire saddlepoint
approach we use will not be applicable and we have no theoretical
tools to establish the rate from Euclidean methods (and probably
real-time methods will also not be applicable since they also rely on
a scale separation between the size and evolution time scale of
sphalerons and the intrinsic thermal size and time scales).

\subsection{Calculation without gradient flow}

We are interested in a configuration close to some point on a
separatrix surface, and in the degree of freedom orthogonal to the
separatrix.  We will write this degree of freedom as $\phi(t)$ with
$\phi=0$ representing the separatrix, and we
take it to have a canonically normalized kinetic term (recall that a
rescaling of $\phi(t)$ does not affect our answers).
For the toy model, $\phi(t) = \sqrt{m}(\varphi(t)-\pi)$.
The probability that the Euclidean fluctuations carry $\phi$ across
$\phi=0$ between $t=0$ and $t=\beta/2$ are given by:
\begin{equation} \label{eq:path_integral}
  \GammaE \equiv \int_{\phi(\tE=0) = \phi(\tE=\beta)}
  \DD\phi \; e^{-S_E} \;
  \left[ \vphantom{\Big|}
    \theta(\phi(\tE{=}0)) \theta(-\phi(\tE{=}\beta/2))
    + \theta(-\phi(\tE{=}0)) \theta(\phi(\tE{=}\beta/2)) \right].
\end{equation}
The two $\theta(..)\theta(..)$ factors represent cases where $\phi$
goes from positive to negative and where it goes from negative to
positive, respectively.  Here the Euclidean action $S_E$ is
\begin{equation}
    S_E = \int_0^\beta \dd{\tE} \left( \frac{1}{2}\left(\partial_\tE \phi \right)^2 - \frac{m^2}{2} \phi^2 \right),
\end{equation}
with $m^2$ representing a possible potential, favoring $\phi$ to leave
the separatrix (the curvature in the unstable direction of the saddlepoint).
We insert the following Matsubara decomposition, which satisfies the
periodicity condition but is otherwise general:
\begin{equation}
	\phi(\tE) = \phi_0 + \sum_{n=1}^{\infty}\left( c_n \cos(2\pi n \tE/\beta) + s_n \sin(2 \pi n \tE/\beta) \right).
\end{equation}
With this substitution, the measure of the path integral becomes simply
\begin{equation}
	\DD\phi = \mathcal{N} \dd{\phi_0} \prod_{k=1}^{\infty} \dd{c_k} \dd{s_k}
\end{equation}
where $\mathcal{N}$ is a constant, and we can explicitly evaluate the action
\begin{align}
    \label{eq:action}
    S_E & = - \frac{\beta\, m^2}{2} \phi_0^2
    + \sum_{n=1}^{\infty}(c_n^2+s_n^2) A(n,m) \,,
    \\ \nonumber
    A(n,m) & \equiv\frac{\beta}{4} \left[\left(\frac{2 \pi
        n}{\beta}\right)^2-m^2\right] .
\end{align}
Introducing $\ceven$ and $\codd$ as the sum over the even and odd
$c_k$ coefficients through
\begin{equation}
  1 = \int \dd\ceven \, \dd\codd \;
  \delta\left(\ceven - \sum_{k=2,4,\ldots} c_k \right)
  \delta\left(\codd - \sum_{k=1,3,\ldots} c_k \right)
\end{equation}
we can rewrite the Heaviside theta-functions as:
\begin{equation}
\begin{split}
  \theta(\phi(\tE=0)) &=
  \theta \left( \phi_0 + \sum_k c_k \right)
  = \theta(\phi_0 + \ceven + \codd) \,,
  \\
  \theta(-\phi(\tE=\beta/2)) &=
  \theta \left( - \phi_0 - \sum_k (-1)^k c_k \right)
  = \theta(-\phi_0 + \ceven - \codd) \,.
\end{split}
\end{equation}
so that the path integral we have to perform is given by
\begin{align}
  \GammaE & = 2 \int \dd{\phi_0} \int \dd \ceven \dd \codd
  \prod_{k=1}^{\infty} \dd{c_k} \dd{s_k} \,
  \exp(\frac{m^2}{2} \phi_0^2 \beta - \sum_{n=1}^\infty (c_n^2 +
  s_n^2) A(n, m))
   \\ \nonumber & {} \times 
  \delta\left( \ceven{-}\!\! \sum_{k=2,4,\ldots} \!\!c_k \right)
  \delta\left( \codd{-} \!\! \sum_{k=1,3,\ldots} \!\! c_k \right)
  \theta ( \phi_0 {+} \ceven {+} \codd )
  \theta ( -\phi_0 {+} \ceven {-} \codd ).
\end{align}
The factor of 2 accounts for the other $\theta$-function combination,
which gives precisely the same answer as the one shown.

Exponentiating the delta functions
\begin{equation} \label{DDeltas}
  \delta \left(\ceven - \sum_k c_k \right)
  = \int \frac{\dd{\lambda}}{2\pi} \;
  \exp\left( i \lambda \ceven -i \sum_k \lambda c_k \right)
\end{equation}
makes all integrals either Gaussians or modified
Gaussians, and it is straightforward to evaluate them.

This calculation suffers from an unknown overall scaling, but recall
that we are interested in the ratio of this result to the standard
result, which is $\langle |\dd\phi/\dd\tm| \rangle$ times the probability
density for $\phi_0=0$.
Equipartition tells us that
$\langle |\dd\phi/\dd\tm|\rangle = \sqrt\frac{2T}{\pi}$, see
\Eq{equipartition}. 
To evaluate the probability density, we compute the same path
integral, but with the two Heaviside functions replaced by a single
delta function $\delta(\phi_0)$ -- essentially, this determines the
normalization factor $\mathcal{N}$:
\begin{equation}
\mathcal{N} =
\int_{\phi(\tE = 0) = \phi(\tE=\beta)} \DD\phi
  \exp(-\int_0^\beta \dd{\tE} \frac{1}{2}\left(\partial_\tE \phi
  \right)^2) \delta(\phi_0) \,,
\end{equation}
which means that
\begin{equation}
    \mathcal{N} = \prod_{n=1}^\infty \frac{A(n,0)}{\pi}.
\end{equation}
With this, by defining
\begin{equation}
    \begin{dcases}
        S_\text{odd (even)} &\equiv \sum_{n \, \text{odd (even)}} \frac{1}{4 A(n,m)}\\
        C &\equiv \frac{1}{4 S_\text{even}} - \frac{m^2 \beta}{2}\\
        b &\equiv \frac{m^2 \beta}{2} \left(1 + \frac{m^2 \beta}{8C} \right)
    \end{dcases}
\end{equation}
the final result for the integral \Eq{eq:path_integral} is
\begin{equation}
    \label{eq:sph_rate_result}
\GammaE = e^{-\frac{m^2 \beta}{2}} \, \frac{m \beta}{2 \sin(m\beta/2)} \sqrt{\frac{1}{\pi
    S_\text{even} b C}} \arctanh \left( 2 \sqrt{S_\text{odd} b}
\right) \,.
\end{equation}

When the saddlepoint is a strongly semiclassical configuration, the
action should not change too quickly as we move away from it, meaning
that $m/2\pi T \ll 1$. If we expand the result in this limit, it
simplifies:
\begin{equation}
  \GammaE = 4 e^{-m^2\beta/2}\, \sqrt{\frac{S_\text{odd}}{\pi}}
  + \mathcal{O}( \sqrt{\beta} (m\beta)^2)
  = \sqrt{\frac{\beta}{2\pi}} + \OO( \sqrt{\beta} (m\beta)^2).
\end{equation}
This is the rate divided by the probability density to be on the
separatrix.  To fully normalize it by the real-time rate, we should
divide by $\sqrt{2T/\pi}$ which we found above, to find:
\begin{align}
  \nonumber
  & \frac{\GammaE}{\mbox{real-time rate}}
  = \sqrt{\frac{1}{2\pi T}} \left( \sqrt{\frac{2T}{\pi}} \right)^{-1}
  = \frac{1}{2T} \,,
  && \mbox{or} \\
  & \boxed{\mbox{Real-time rate} = 2T \; \GammaE \,.\; }
  \label{Convert}
\end{align}
This provides the desired relation between the Affleck real-time rate
and the Euclidean sphaleron density which we have defined.

\subsection{Calculation with flow}
\label{sec:toy-model-gf}

In practice it is necessary to modify the $\phi(t)$ field inside the
Euclidean path integral through the application of gradient flow,
before performing any evaluations on it.
How does this affect the probability that $\phi(t=0)$ and
$\phi(t=\beta/2)$ are on opposite sides of the separatrix?
We will answer this now.
Gradient flow changes the measurables, not the path integral or path
integral variables.  In our case, it damps the $t$-dependent
fluctuations through application of the heat equation.  Define the
depth of heat-equation flow to be $\tauf$.  Then we set the $\tauf=0$
boundary conditions to be $\phi(\tauf=0,t) = \phi(t)$, and the
gradient flow itself is described by
\begin{equation}
    \partial_\tauf \phi(\tauf, \tE) - \partial_\tE^2 \phi(\tauf, \tE) = 0.
\end{equation}
Using the same Fourier-series expansion for $\phi(t)$ as before, one
easily finds that
\begin{equation}
  \phi(\tauf, t) = \phi_0
  + \sum_{n=1}^{\infty}\left( c_n \cos(2\pi n \tE/\beta)
  + s_n \sin(2 \pi n \tE/\beta) \right)
  \exp(-\tauf \left(2 \pi n/\beta\right)^2).
\end{equation}

Now we can proceed as we did in the unflowed case, but taking the flow
into account on our theta functions, while the action remains the same
as in \Eq{eq:action}.  The Heaviside functions become
\begin{equation}
    \begin{split}
        &\theta(\phi(\tauf, 0)) = \theta \left( \phi_0 + \sum_k c_k \exp(-\tauf \left(2 \pi n/\beta\right)^2) \right) \\
        &\theta(-\phi(\tauf, \beta/2)) = \theta \left( -\phi_0 - \sum_k \left(-1\right)^k c_k \exp(-\tauf \left(2 \pi n/\beta\right)^2) \right).
    \end{split}
\end{equation}
Proceeding as we have done before, and defining in this case
\begin{equation}
    S^\tauf_\text{odd (even)} \equiv \sum_{n \, \text{odd (even)}} \frac{\exp\left(-2\tauf \left(2 \pi n/\beta\right)^2 \right)}{4 A(n,m)}
\end{equation}
we find
\begin{equation}
    \label{eq:sph_rate_flow}
    \begin{split}
        \GammaE^\tauf &= e^{-\frac{m^2}{2} \beta} \frac{m \beta}{2 \sin(m\beta/2)} \sqrt{\frac{1}{\pi S^\tauf_\text{even} b C}} \arctanh \left( 2 \sqrt{S^\tauf_\text{odd} b} \right) \\
        & \Rightarrow \;
        \sqrt{\frac{\beta}{2\pi}}
        \sqrt{\frac{8}{\pi^2}
          \sum_{n=1,3,\ldots} \frac{e^{-8\pi^2 \tauf T^2 n^2}}{n^2}}
        \\ & \simeq
        \sqrt{\frac{\beta}{2\pi}} \times
        \sqrt{1-8 \sqrt{\frac{2\tauf}{\pi \beta^2}}}.
    \end{split}
\end{equation}
As before, the quantity after $\Rightarrow$ is the result when taking
$m\to 0$.  The last line is the result of a small $\tauf$ expansion,
with errors which are exponentially small provided that
$\tauf \ll 1/(8\pi^2 T^2)$.  In practice we will use the unexpanded
expression shown in the middle line.
The final expression is useful because it shows that the correction
goes to 1 as $\tauf \to 0$ but only as a square root, that the
behavior is monotonic, and that the domain over which the gradient
flow can be viewed as ``small'' is roughly
$\tauf \ll 1/(8 \pi^2 T^2)$.

The best way to use this result is:
\begin{equation}
  \boxed{ \mbox{Real-time rate} =
    \frac{2T \; \GammaE^\tauf}{\sqrt{1-8\sqrt{\frac{2\tauf
            T^2}{\pi}}}}\,.\; }
\label{ConvertFlow}
\end{equation}

\bibliographystyle{unsrt}
\bibliography{refs}

\end{document}